\documentclass[iop]{emulateapj}
\usepackage{graphicx}
\usepackage{amssymb}
\usepackage{amsmath}
\usepackage{hyperref}
\usepackage{natbib}
\usepackage{longtable}
\usepackage{color}
\usepackage{float}

\newcommand{\Kepler}{{\it Kepler}}

\newcommand{\kms}{km~s$^{-1}$}
\newcommand{\teff}{$T_\text{eff}$}
\newcommand{\jktebop}{{\sc jktebop}}
\newcommand{\msun}{M$_\odot$}
\newcommand{\rsun}{R$_\odot$}
\newcommand{\lsun}{L$_\odot$}
\newcommand{\mjup}{M$_\text{Jup}$}

\newcommand{\starA}{{EPIC\,203476597}}
\newcommand{\starB}{{EPIC\,203710387}}
\newcommand{\starC}{{EPIC\,203868608}}

\shorttitle{Eclipsing Binaries in Upper Scorpius}
\shortauthors{David et al.}

\begin{document}

\title{K2 Discovery of Young Eclipsing Binaries in Upper Scorpius: \\ Direct Mass and Radius Determinations for the Lowest Mass Stars and Initial Characterization of an Eclipsing Brown Dwarf Binary}

\slugcomment{submitted to ApJ 20 August 2015; revised 10 November 2015}

\author{Trevor J. David \altaffilmark{1,2}, Lynne A. Hillenbrand \altaffilmark{1}, Ann Marie Cody \altaffilmark{3}, John M. Carpenter \altaffilmark{1}, Andrew W. Howard \altaffilmark{4}}

\altaffiltext{1}{Department of Astronomy, California Institute of Technology, Pasadena, CA 91125}
\altaffiltext{2}{NSF Graduate Research Fellow}
\altaffiltext{3}{NASA Ames Research Center, Mountain View, CA 94035}
\altaffiltext{4}{Institute for Astronomy, University of Hawaii, 2680 Woodlawn Drive,
Honolulu, HI 96822, USA}

\email{tjd@astro.caltech.edu}

\begin{abstract}
We report the discovery of three low-mass double-lined eclipsing binaries in the pre-main sequence Upper Scorpius association, revealed by $K2$ photometric monitoring of the region over $\sim$ 78 days. The orbital periods of all three systems are $<$5 days.  We use the $K2$ photometry plus multiple Keck/HIRES radial velocities and spectroscopic flux ratios to determine fundamental stellar parameters for both the primary and secondary components of each system, along with the orbital parameters. We present tentative evidence that EPIC 203868608 is a hierarchical triple system comprised of an eclipsing pair of $\sim$25 $M_\mathrm{Jup}$ brown dwarfs with a wide M-type companion. If confirmed, it would constitute only the second double-lined eclipsing brown dwarf binary system discovered to date. The double-lined system EPIC 203710387 is composed of nearly identical M4.5-M5 stars with fundamentally determined masses and radii measured to better than 3\% precision  ($M_1=0.1183\pm0.0028 M_\odot$, $M_2=0.1076\pm0.0031 M_\odot$ and $R_1=0.417\pm0.010 R_\odot$, $R_2=0.450\pm0.012 R_\odot$) from combination of the light curve and radial velocity time series. These stars have the lowest masses of any stellar mass double-lined eclipsing binary to date. Comparing our derived stellar parameters with evolutionary models suggest an age of $\sim$10-11 Myr for this system, in contrast to the canonical age of 3-5 Myr for the association. Finally, EPIC 203476597 is a compact single-lined system with a G8-K0 primary and a likely mid-K secondary whose line are revealed in spectral ratios. Continued measurement of radial velocities and spectroscopic flux ratios will better constrain fundamental parameters and should elevate the objects to benchmark status. We also present revised parameters for the double-lined eclipsing binary UScoCTIO 5 ($M_1=0.3336\pm0.0022 M_\odot$, $M_2=0.3200\pm0.0022 M_\odot$ and $R_1=0.862\pm0.012$, $R_2=0.852\pm0.013 R_\odot$), which are suggestive of a system age younger than previously reported. We discuss the implications of our results on these $\sim$0.1-1.5 $M_\odot$ stars for pre-main-sequence evolutionary models.
\end{abstract}

\section{Introduction}
\label{sec:intro}

In the aftermath of the loss of two of its reaction wheels, the \Kepler\ spacecraft was reoriented to observe fields along the ecliptic plane for consecutive campaigns of $\sim$75 days in duration and designated as the $K2$ extended mission \citep{howell2014}.  The $K2$ Field 2 pointing encompasses the Upper Scorpius region of recent star formation  \citep[see][for a review]{pm2008} and the molecular cloud near $\rho$ Ophiuchus in which star formation is ongoing \citep[see][for a review]{wilking2008}. 

Extinction is quite high towards the ``$\rho$ Oph" molecular cloud, but some cluster members (typically those of higher mass) are bright enough for study with $K2$. The sizable ``Upper Sco" association by contrast is essentially gas free, though there is a small amount of dust extinction ($A_V < 1$). The association samples a wide range in mass -- from mid-B type stars having several to ten solar masses, all the way down to late M-type, very low mass stars and sub-stellar mass objects, the majority of which are bright enough for $K2$ photometry. 

Census work in the Upper Sco region has established over 1500 secure and candidate members, with a major compilation of candidates appearing in \cite{lodieu2013}. Notable studies include the early kinematic work that culminated in \cite{preibisch2002} as well as contemporaneous x-ray \citep[e.g.][]{kohler2000} and wide-field optical \citep[e.g.][]{ardila2000} studies, through to the most recent additions to the stellar population by e.g. \cite{rizzuto2011,rizzuto2015} and \cite{gagne2015}. The traditional age of the association is 3-5 Myr  \citep[e.g.][]{degeus1989, preibisch2002, slesnick2008}  which is reinforced in the analysis of \cite{hh2015}, hereafter HH15, using modern pre-main sequence tracks and a sample of several hundred GKM stars, but challenged by \cite{pecaut2012} who argue for an age of 11 Myr based on an assessment of  5-6 post-main sequence stars and several tens of AFG stars near the main sequence.  The $\rho$ Oph region is significantly younger at $<$1-2 Myr and features self-embedded protostars, classical T Tauri disks in various stages of evolution, and disk-free young stars; \cite{wilking2008} provide a compilation of accepted members.

Notably, in the short but exciting time span between the age of younger active star-forming regions such as $\rho$ Oph and the only somewhat older Upper Sco region, definitive changes are taking place in both the stars and their circumstellar environments. Most relevant for this paper is that the stars will have contracted by a mass-dependent factor of 50\% to 250\%, making the existing $K2$ data a valuable resource for measuring the pre-main sequence evolution of stellar radii.

Pre-main-sequence eclipsing binaries (EBs) are particularly valuable for calibrating pre-main sequence evolutionary models which show large discrepancies when compared with EB measurements and remain poorly constrained at the very lowest masses ($M < 0.3$ \msun), as reviewed by \cite{stassun2014}. Only a small number of such pre-main sequence EB systems are known. Here, we report the discovery of three new pre-main sequence EBs, two secure members and one likely member of Upper Sco, and all with short periods ($<$ 5 days).

In \S~\ref{sec:photometry} we discuss characteristics of the $K2$ observations as well as our procedures for light curve extraction and subsequent removal of intrinsic and systematic variability. We discuss our spectroscopic observations, which we use to measure radial velocities, establish spectral types, and confirm membership, in \S~\ref{sec:spectra}. The procedures for determination of orbital and stellar parameters are described in \S~\ref{sec:orbitfitting} and \S~\ref{sec:stellarparams}, respectively. Finally, we discuss the individual EB systems and our results on fundamentally determined radii and masses in \S~\ref{sec:individual}.

\section{$K2$ Observations and Analysis}
\label{sec:photometry}

\begin{figure}
\centering
\includegraphics[width=0.475\textwidth]{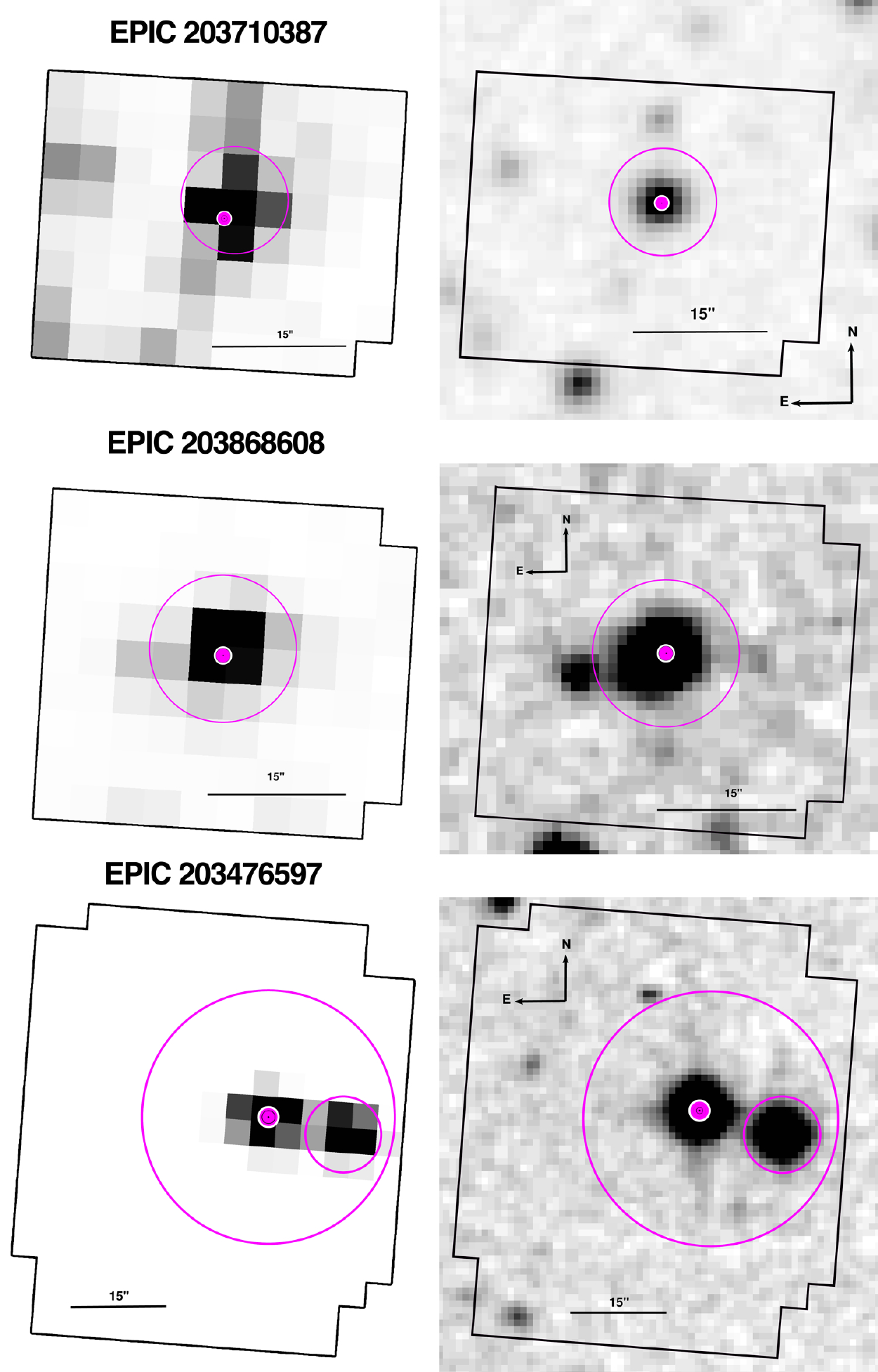}
\caption{\emph{Left column:} $K2$ postage stamps showing the regions around the three EB systems.  Orientation is such that north is up and east is left. The $K2$ plate scale is $\sim$ 4\arcsec\ /pixel.  The magenta circles indicate the photometric apertures used for light curve extraction. The points represent the nominal locations of the sources from the target pixel file header information and may not be centered on the star due to small errors in the WCS (World Coordinate System). For \starA, the second, smaller aperture around the neighboring star to the west was used to compute the time-averaged flux which was ultimately subtracted from the raw EB light curve. \emph{Right column:} DSS2 ``infrared" views of the corresponding regions presented on the left.   The potential for contamination in the $K2$ photometry due to either unresolved or spatially resolved nearby sources is discussed in the text individually for each EB system.
} 
\label{fig:stamp}
\end{figure}

A field covering the Upper Sco region was observed by $K2$ in 2014 between BJD 2456894 - 2456970. The modified observing configuration of $K2$ has the telescope pitch and yaw confined using the two remaining reaction wheels, while the roll along the boresight is partially balanced by solar radiation pressure with thruster firings every $\sim$6 hours to correct for the remaining azimuthal drift. As a result of the roll axis drift and intrinsic flat field variations, $K2$ light curves possess significant systematic noise correlated with the telescope pointing. After correcting for this systematic noise, the photometric precision of $K2$ light curves over typical transit timescales of $\sim$6 hours has been measured to be a magnitude-dependent factor of 2-3 lower than that of \Kepler\ data \citep{vanderburg2014b, aigrain2015}. 

The precision of \Kepler\ light curves can be quantified by the metric of Combined Differential Photometric Precision (CDPP) originally described by \cite{christiansen2012}. We use a quasi-CDPP, defined as the median of the standard deviation in a running bin of a fixed duration. For this work, we choose 6.5-hour as the time frame over which to calculate the quasi-CDPP which is used as our light curve precision metric.

\subsection{Light Curve Extraction}
\label{subsec:lcextraction}

In $K2$ Field 2, we extracted photometry for objects identified as being members or candidate members of Upper Sco and the slightly younger $\rho$ Ophiuchus complex, which is nearby and somewhat overlapping in projection on the sky. Aperture photometry was performed with the {\sc Python} \texttt{photutils} package on background-subtracted images using a range of aperture radii from 1.5 to 5 pixels. Unlike the \Kepler\ pipeline and recently publicized reductions of $K2$ data \citep[e.g.][]{vanderburg2014b, foremanmackey2015}, we vary aperture placement with the stellar centroid position. We computed a flux-weighted centroid in a 7$\times$7 pixel box centered on the location of each star, as specified by the target pixel file header information. Stellar flux within the aperture was computed using the \texttt{photutils} ``exact'' setting, in which the intersection of the circular aperture with the square pixels is calculated. The $K2$ regions used for aperture photometry, and corresponding ``infrared" views of the regions from DSS2\footnote{http://archive.stsci.edu/dss/acknowledging.html}, are shown in Figure~\ref{fig:stamp}.

Depending on detector position, we find that source centroids move at up to 0.03 pixel (i.e., 0.12\arcsec) per hour due to instability of the telescope's pointing. Approximately every six hours, a correction is applied to return pointing to the nominal position. Since intrapixel sensitivity can vary at the few percent level, even small centroid movements can contribute systematic effects to $K2$ photometry. Shifting the aperture according to centroid position partially mitigates these effects. For light curves with significant pointing related systematics, we also applied a detrending procedure to recover the intrinsic variability pattern (see \S~\ref{subsec:detrending}).

For many stars, signatures are present in the raw photometric extractions of behavior associated with e.g. young star accretion or circumstellar obscuration, starspots and stellar rotation, chromospheric flaring, and binary eclipses. However, both the light curves with these types of large amplitude variations, and those light curves with more subtle variations, can benefit from attention to so-called de-trending which aims to remove prominent systematic effects and restore the innate photometric precision of the \Kepler\ spacecraft CCDs.

\subsection{Detrending Procedure}
\label{subsec:detrending}

Multiple techniques for detrending $K2$ light curves have emerged in the literature \citep[e.g.][]{vanderburg2014b, aigrain2015, lund2015, huang2015}. \cite{foremanmackey2015} advocate fitting systematic effects simultaneously with the astrophysical signals sought to be quantified (e.g. transits). This approach, as those authors point out, mitigates the risk of distorting the astrophysical signal in question through under- or over-fitting. 

\begin{figure*}
\centering
\includegraphics[width=0.88\textwidth]{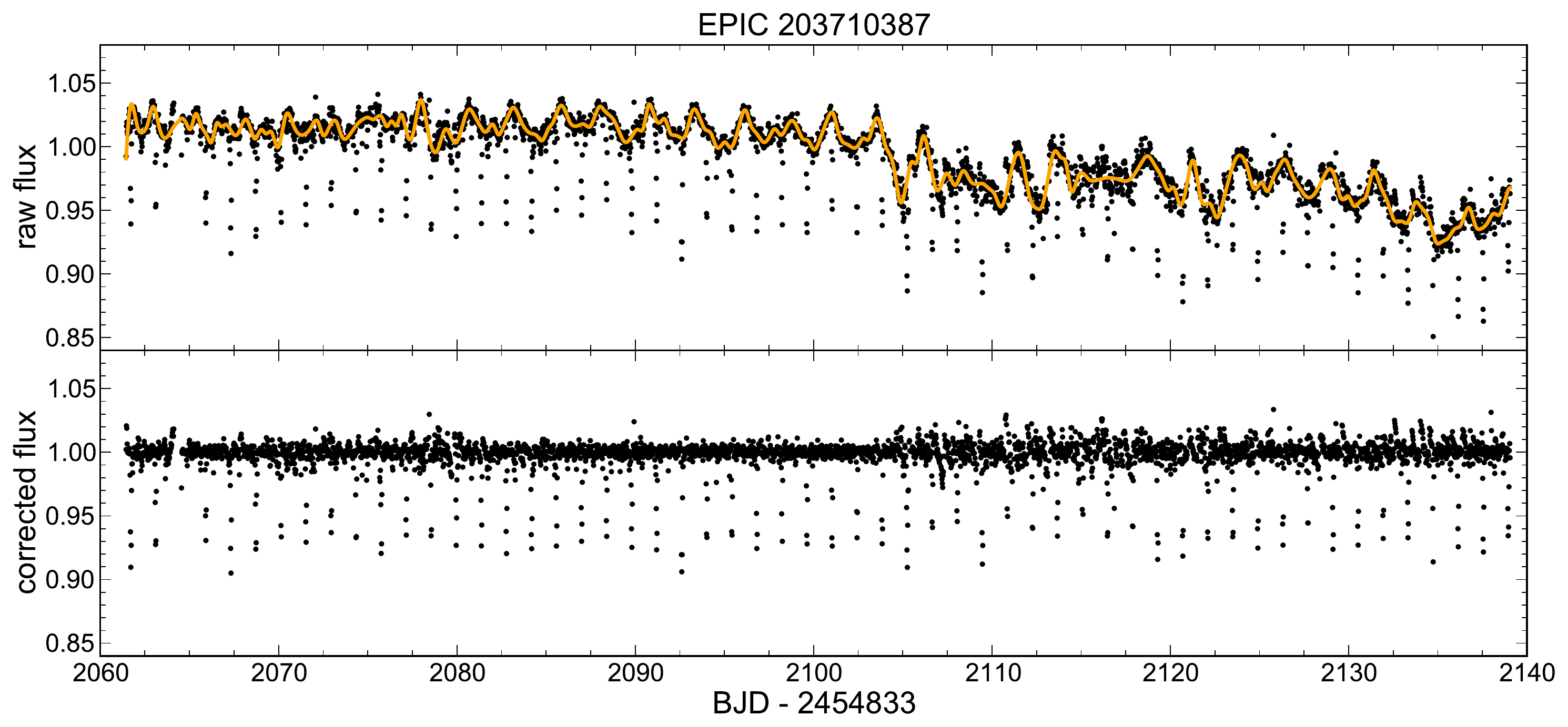}
\includegraphics[width=0.88\textwidth]{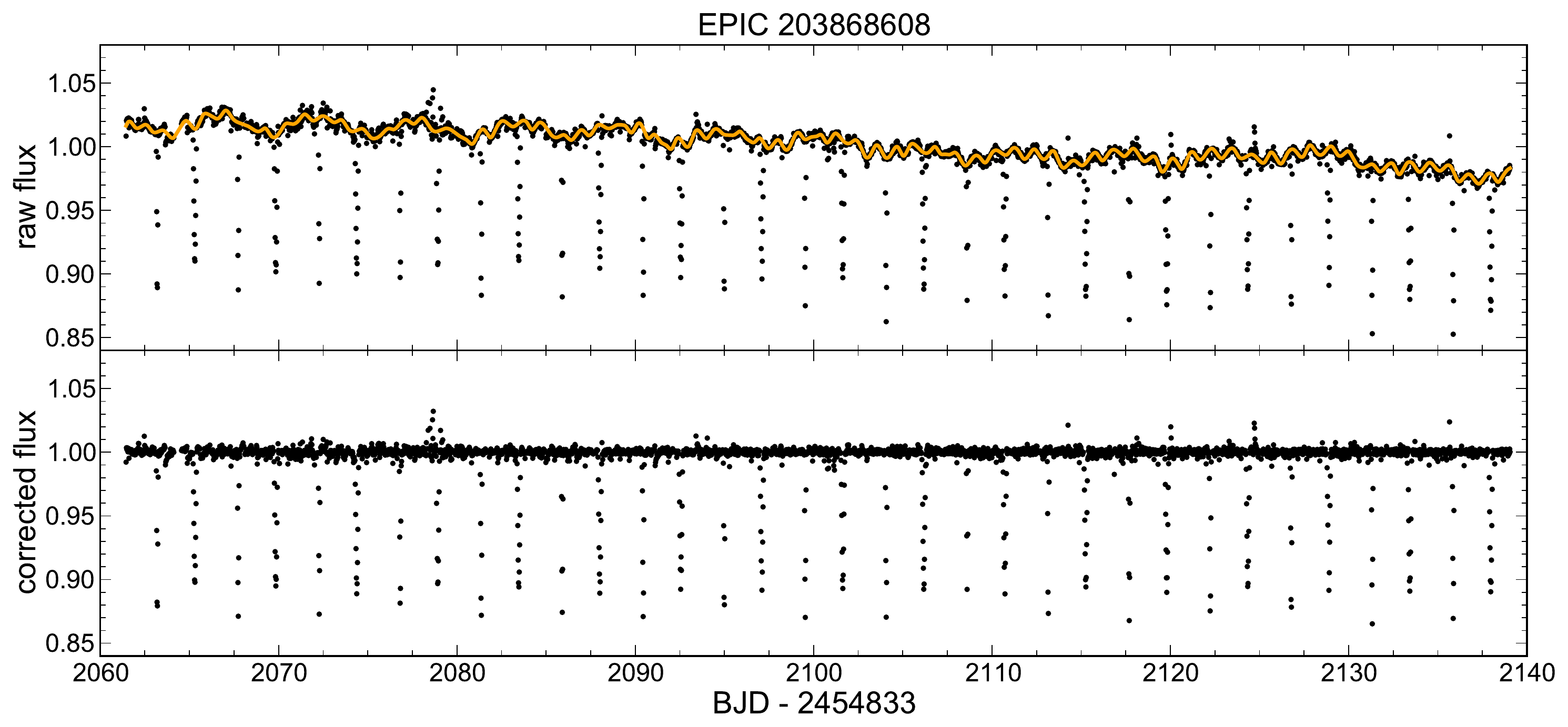}
\includegraphics[width=0.88\textwidth]{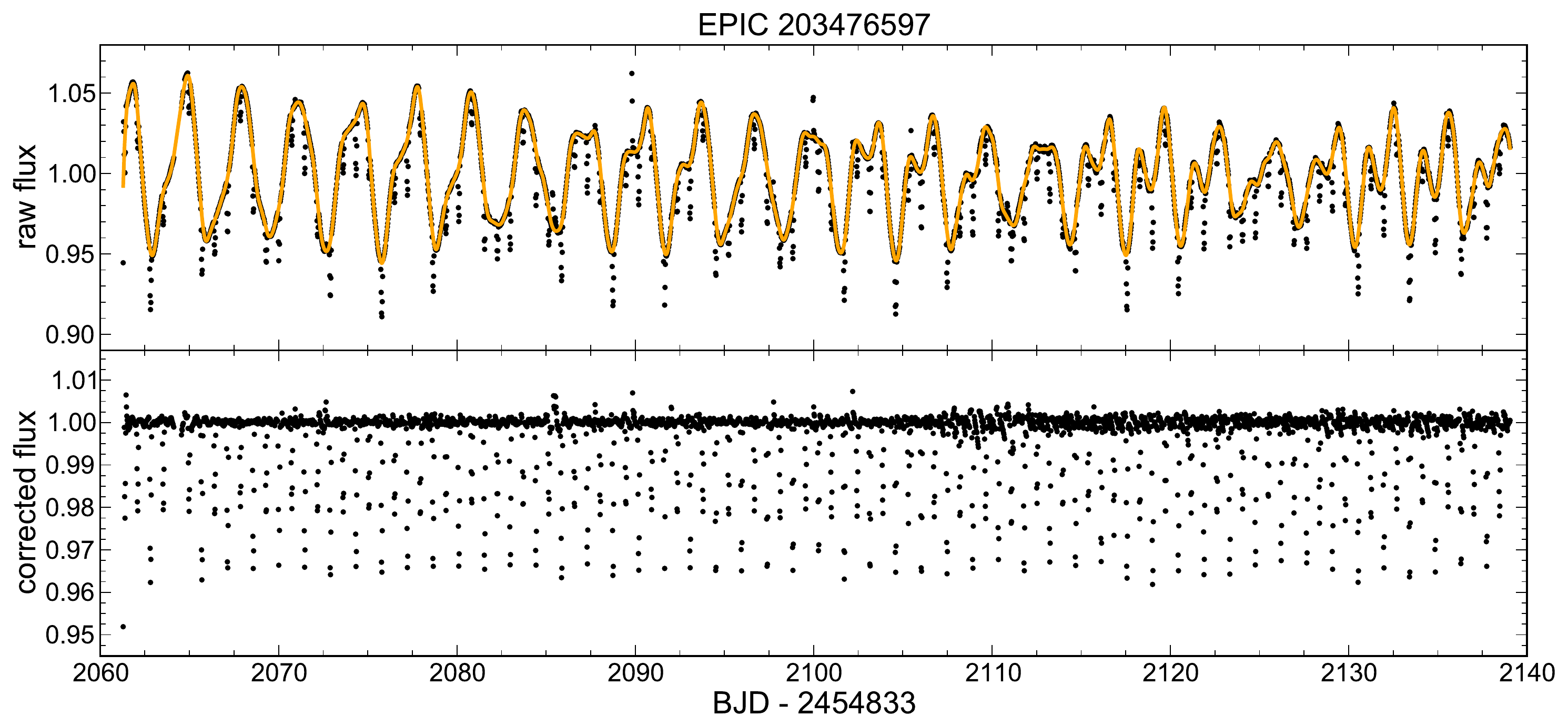}
\caption{For each of the three eclipsing binary systems, raw $K2$ (top panels) and corrected (bottom panels) light curves. All fluxes are median normalized. The orange line indicates the cubic B-spline fits to the raw photometry used to produce the corrected fluxes. A noticeable change in the data quality between the first and second halves of Campaign 2 is seen in these sources (most prominently \starB) as well as many others.}
\label{fig:lightcurve}
\end{figure*}

In this work, the raw light curves are corrected for systematic and astrophysical variability through a principal component analysis procedure based on that of \cite{vanderburg2014} and \cite{vanderburg2014b}, hereafter V14. The approach employed here differs from that of V14 in the following ways:

\begin{enumerate}
\item{We opted to detrend data from the entire campaign at once, as opposed to dividing the campaign into smaller sets of observations.}
\item{We removed outlier points with nonzero quality flags, corresponding to e.g. attitude tweaks and observations taken in coarse pointing mode, except those observations with the detector anomaly flag raised as these are fairly common. We also discarded any observations that were simultaneous 3-$\sigma$ outliers in both $x$ and $y$ centroid coordinates.}
\item{We considered photometry generated from four possible apertures of radii 1.5, 2, 3 and 5 pixels, selecting the raw light curve with the lowest 6.5 hr quasi-CDPP.}
\item{Principle component analysis is used to transform the $x,y$ centroid positions to a new coordinate space, $x',y'$, in which the positions drift primarily along one axis. A polynomial fit to the new $x',y'$ coordinates is then performed in order to determine the `arclength' (defined in V14) at each position. Instead of only a degree 5 fit to the transformed coordinates, we perform polynomial fits of degrees 1 through 5, and select the best-fitting curve (after ten iterations of 3-$\sigma$ outlier exclusion) according to a Spearman test.}
\item{The raw photometry is corrected for the centroid position variability effects via the process above, which produces a ``low-pass filtered" flux (i.e. corrected for trends on timescales $<$6 hours). 

In some instances, this step of the detrending procedure can introduce additional noise to the raw photometry (as was the case for the three EBs discussed here). This is partially due to the fact that we detrend the entire campaign of data at once, and the pointing-related trends are often of shorter duration (on the order of days).  It is also likely that allowing the photometric aperture to shift with centroid position, as we do, partially mitigates pointing-induced trends.  Thus, at this stage, the 6.5 hour quasi-CDPP of the raw photometry is compared with that of the low-pass flux and the higher quality light curve is selected for ``long-term" variability correction.}
\item{
As a final step, we correct for variability on timescales longer than 6 hours. The source of variability on these timescales can be a combination of astrophysical (as is the case with \starA, seen in Figure~\ref{fig:lightcurve}), pointing-related effects, and long-term systematic trends (such as a general decline in overall flux levels seen from the first to second halves of the campaign). V14 correct for long-term systematic variability via an iterative spline fit, with knots every 1.5 days and 3-sigma outlier rejection to ensure that transit signals do not drag down the spline fit thus resulting in distorted transit signals in the corrected light curve. In our iterative intrinsic variability fitting, we allow a much more flexible spline with knots every 12 cadences ($\sim$6 hours) and up to 10 iterations with 2-$\sigma$ outlier rejection at each stage. This approach appropriately fits and removes the intrinsic variability exhibited by these young stars, which can be significant over short timescales similar to the timescales expected for eclipse/transit durations. Our aggressive approach to outlier rejection ensures that any eclipse/transit signals are excluded from the variability fit. The \texttt{splrep} and \texttt{splev} tasks in the \texttt{scipy.interpolate} package were used to perform the spline fit in {\sc Python}.}
\end{enumerate}

The raw $K2$ and corrected light curves for each of the EB systems are depicted in Figure~\ref{fig:lightcurve}.

\section{Spectroscopic Observations and Analysis}
\label{sec:spectra}

\begin{figure*}
\centering
\includegraphics[width=0.74\textwidth,angle=-90]{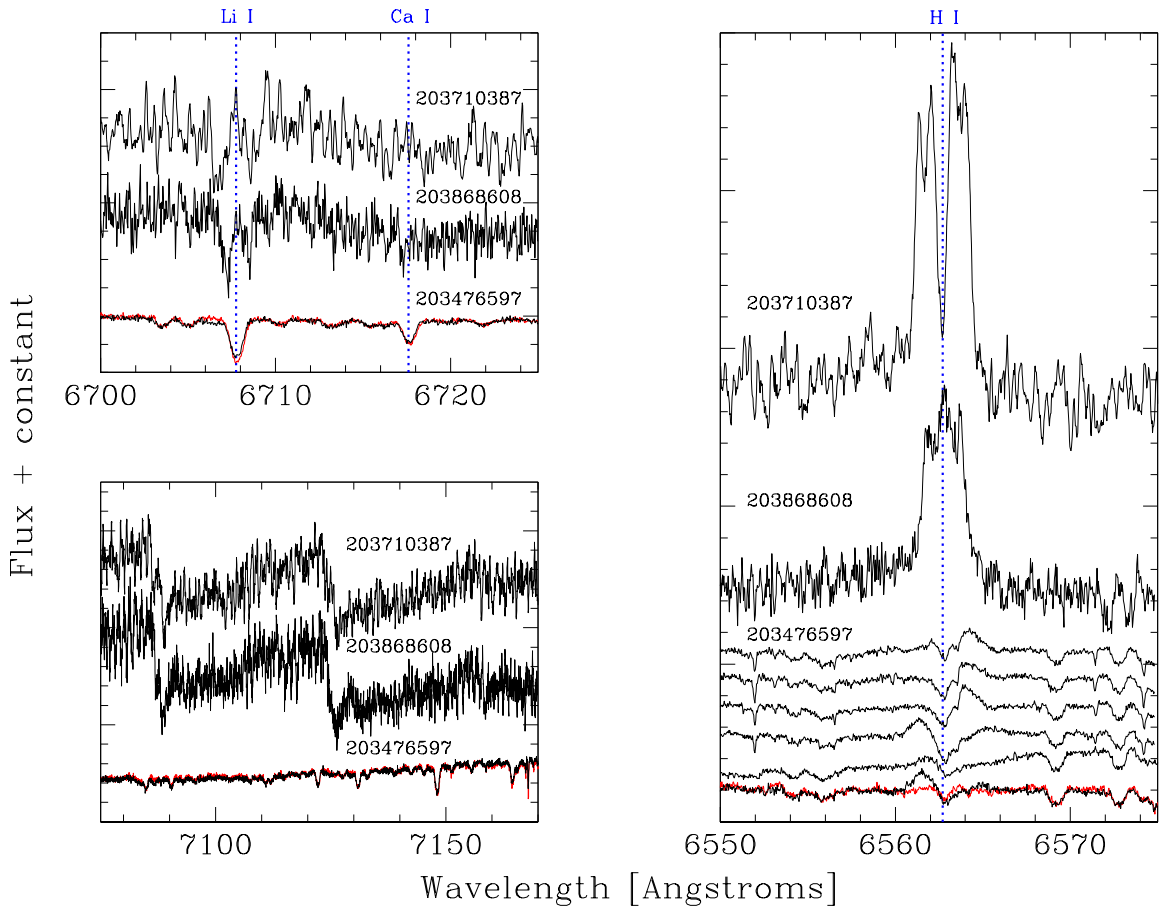}
\caption{Sections of the HIRES spectra showing a photospheric region (lower left), the \ion{Li}{1} 6707.8 \AA\ and \ion{Ca}{1} 6717 \AA\ lines (upper left) and the H$\alpha$ line profiles (right). All three stars show H$\alpha$ activity and have \ion{Li}{1} absorption. The spectra of \starB\ and \starC\ are clearly double-lined, as seen most prominently in the two components of H$\alpha$ emission (each of which is double-peaked) and in the doubled \ion{Li}{1} absorption, but also in the TiO bandhead regions. The red line indicates a second spectrum of \starA, which differs from the black (first) that it overlays in its H$\alpha$ profile and in the \ion{Li}{1} line, where a small absorption blueward of line center moves to become enhanced absorption redward of line center. The full time series of spectra is shown for \starA\ in the H$\alpha$ panel; we interpret the profile variations as due to orbital motion of a faint young H$\alpha$ emitting secondary, which is indeed revealed in the absorption lines from differences and ratios of the spectra.
}
\label{fig:spectra}
\end{figure*}

We obtained initial high dispersion spectra for the three EBs on 1 and 2 June 2015, UT using Keck I and HIRES \citep{vogt1994}. The instrument was configured to produce spectra from $\sim$4800-9200 \AA\ using the C5 decker which provides spectral resolution $\sim$36,000. Additional HIRES spectra were obtained using the setup of the California Planet Search covering $\sim$3600-8000 \AA\ at R$\sim$48,000 with the C2 decker, on the six additional nights listed in Table \ref{table:rvs}. Figure~\ref{fig:spectra} shows for all three stars a photospheric region of spectrum along with the profiles of H$\alpha$ and \ion{Li}{1} 6707.8 \AA.  

We use the spectra to assess spectral types, to confirm membership through detection of H$\alpha$ emission and \ion{Li}{1} absorption, and to measure systemic radial velocities from binary orbit fitting. The equivalent widths are given in Table~\ref{table:ews}; line strengths are consistent with the expectations for young active low mass stars with some variation observed among the epochs in the H$\alpha$ strengths.  We note that our measurements for \starA\ match within expectations the values reported by \cite{rizzuto2015} from lower resolution spectra. 

The FXCOR task within IRAF\footnote{IRAF is distributed by the National Optical Astronomy Observatory, which is operated by the Association of Universities for Research in Astronomy (AURA) under a cooperative agreement with the National Science Foundation} was used to measure relative velocities, using selected spectral orders with sufficient S/N, and spectral ranges with abundant photospheric features and minimal atmospheric contamination. FXCOR implements the \cite{tonry1979} method of cross correlation peak finding; a Gaussian profile was used to interactively fit for the velocity shift and errors for individual components of each binary at each epoch. The measured velocities were calibrated to radial velocity standard stars as detailed below, with each spectrum first corrected to the heliocentric frame. The final velocities at each epoch are derived as weighted means from among the individual orders.
The results on radial velocities are discussed below in the sections on the individual EB systems, and are presented in Table~\ref{table:rvs}.

At any given epoch, the relative heights of the cross correlation peaks for the two components of a double-lined binary system can be used as an approximation of the flux ratio, with final values again taken as means among the measured orders.

\begin{deluxetable*}{ccrcrcc} 
\tabletypesize{\footnotesize} 
\tablewidth{0.95\textwidth} 
\tablecaption{ Keck-I/HIRES Radial Velocities and Flux Ratios \label{table:rvs}} 
\tablehead{ 
\colhead{EPIC} & \colhead{Epoch} & \colhead{$v_1$} & \colhead{$\sigma_{v_1}$} & \colhead{$v_2$} & \colhead{$\sigma_{v_2}$} & \colhead{$F_2/F_1$} \\
\colhead{identifier} & \colhead{(BJD-2450000)} & \colhead{(\kms)} & \colhead{(\kms)} & \colhead{(\kms)} & \colhead{(\kms)} &
}
\startdata 
203710387 & 7174.83185 &  38.98 $\pm$ 0.31 & 0.69 & -51.31 $\pm$ 0.37 & 0.86 & 0.939 $\pm$ 0.175 \\
...        & 7175.83303 & -26.03 $\pm$ 0.45 & 0.69 &  19.40 $\pm$ 0.44 & 1.04 & 0.976 $\pm$ 0.139 \\
...        & 7176.02710 & -38.94 $\pm$ 0.39 & 1.27 &  35.29 $\pm$ 0.38 & 1.13 & 0.937 $\pm$ 0.198 \\ 
...        & 7217.80815 & -10.14 $\pm$ 0.20 & 0.35 &   5.87 $\pm$ 0.25 & 0.91 & 0.947 $\pm$ 0.055 \\
...        & 7254.84470 & -45.77 $\pm$ 0.07 & 1.01 &  42.36 $\pm$ 0.08 & 0.93 & 0.951 $\pm$ 0.093 \\
...        & 7255.82026 &  14.23 $\pm$ 0.14 & 0.90 & -21.49 $\pm$ 0.14 & 1.26 & 0.971 $\pm$ 0.057 \\
203868608 & 7175.92133 &  -5.24 $\pm$ 0.12 & 0.42 & ... & ... & ... \\
...        & 7217.81681 &  16.51 $\pm$ 0.01 & 0.25 & -29.5 $\pm$ 0.01 & 0.47 & 0.980 $\pm$ 0.037 \\ 
...        & 7255.82988 & -26.39 $\pm$ 0.03 & 1.62 & 14.51 $\pm$ 0.03 & 0.51 & 1.075 $\pm$ 0.056 \\
...        & 7262.79913 &  21.87 $\pm$ 0.04 & 0.80 & -25.48 $\pm$ 0.06 & 1.19 & 1.120 $\pm$ 0.100 \\ 
...        & 7265.79721 &  -4.66 $\pm$ 0.02 & 0.23 & ... & ... & ... \\
...        & 7290.72899 &  15.79 $\pm$ 0.03 & 0.19 & -29.26 $\pm$ 0.03 & 0.26 & 0.955 $\pm$ 0.071 \\
203476597 & 7175.84692 & -1.66 $\pm$ 0.33 & 0.67 & ... & ... & ... \\
      ... & 7176.05433 & -0.12 $\pm$ 0.29 & 0.95 & ... & ... & ...\\
      ... & 7217.82236 & -0.72 $\pm$ 0.19 & 0.56 & ... & ... & ... \\
      ... & 7254.83534 & -0.44 $\pm$ 0.11 & 0.51 & ... & ... & ... \\
      ... & 7255.81131 &  0.02 $\pm$ 0.12 & 0.38 & ... & ... & ... \\
      ... & 7262.79242 & -0.47 $\pm$ 0.13 & 0.64 & ... & ... & ... \\
      ... & 7265.79154 & -1.29 $\pm$ 0.13 & 0.42 & ... & ... & ...
\enddata 
\tablecomments{Quoted radial velocities are weighted means across several spectral orders within a single epoch, with each measurement weighted inversely to the variance. Formal errors on the weighted mean are quoted to the right of each measurement, where the errors are defined as the square root of the variance of the weighted mean (defined as $\sigma^2 = 1/\sum^{n}_{i=1} \sigma_i^{-2}$). The uncertainties actually used in the orbital parameter fitting procedure, $\sigma_v$, are the root-mean-square errors between individual measurements. The final column lists flux ratios, measured from the relative peak heights in the cross-correlation functions of double-lined systems.}
\end{deluxetable*}

\begin{deluxetable}{ccc} 
\tabletypesize{\footnotesize} 
\tablewidth{0.45\textwidth} 
\tablecaption{ Keck-I/HIRES Equivalent Widths \label{table:ews}} 
\tablehead{ 
\colhead{EPIC} & \colhead{EW(H$\alpha$)} & \colhead{EW(\ion{Li}{1} 6707.8)} \\
\colhead{identifier} & \colhead{(\AA)} & \colhead{(m\AA)} 
}
\startdata 
203710387A & -2.9 & 150 \\
203710387B & -2.4 & 420 \\
203868608A & -1.8 & 260 \\
203868608B & -1.4 & 310 \\  
203476597A & weak abs. & 360 \\
203476597B & -0.2: & 95:    
\enddata 
\tablecomments{
Numbers correspond to the spectra shown in Figure~\ref{fig:spectra}.
The H$\alpha$ measurements have $\sim$0.1 \AA\ measurement accuracy 
but up to 30\% variation among epochs, and the \ion{Li}{1} measurement error is estimated at $<$5-10\%.}
\end{deluxetable}

\section{Orbital Parameter Fitting}
\label{sec:orbitfitting}

\begin{figure*}
\centering
\includegraphics[width=0.99\textwidth]{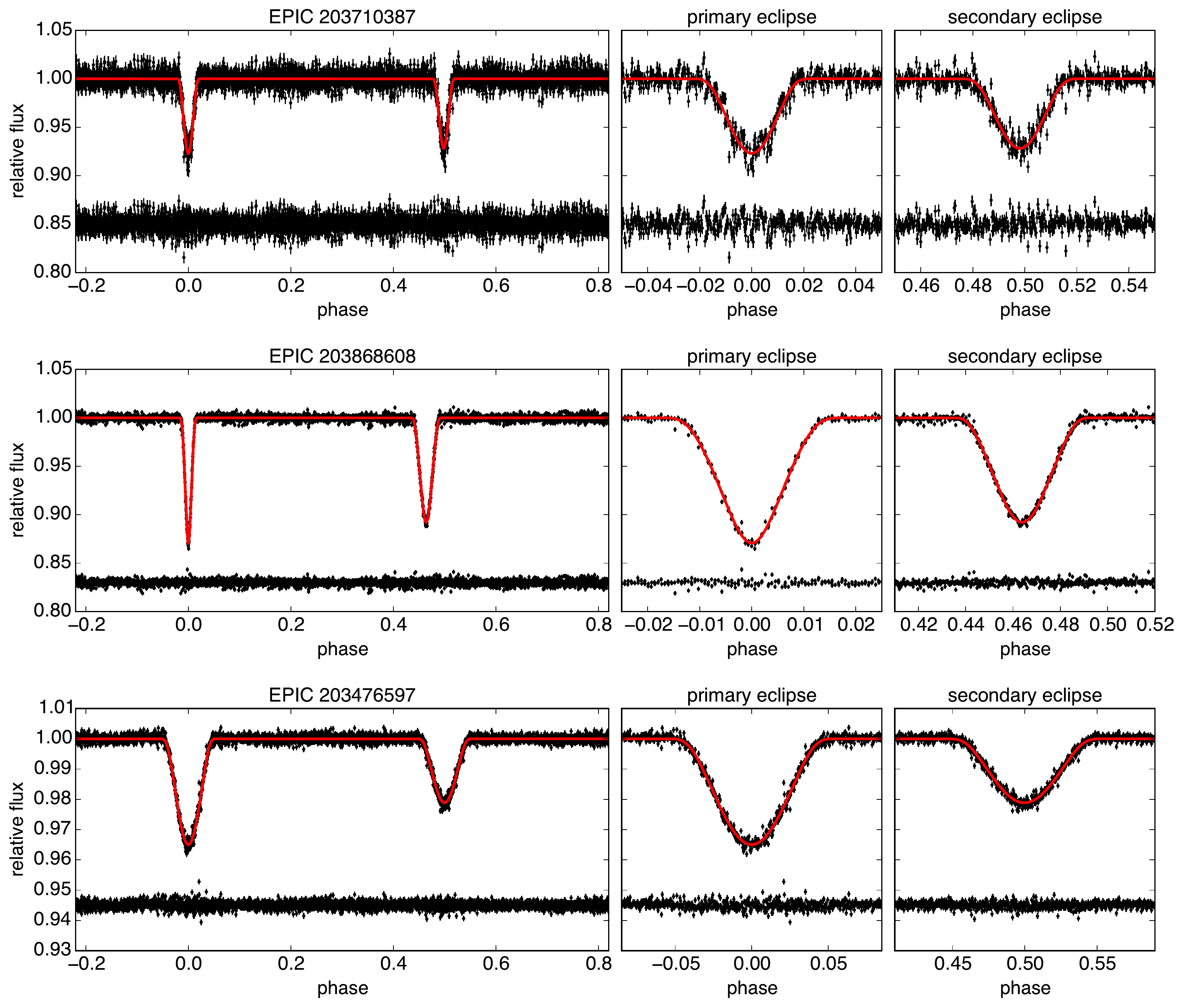}
\caption{Phased $K2$ light curves (black points) with best-fitting \jktebop\ models (red curves). Residuals are plotted below the model fits. Observational errors are determined by the RMS scatter in the out-of-eclipse portions of the light curves. From top to bottom, the periods of these three EBs are approximately 2.8 d, 4.5 d, and 1.4 d.}
\label{fig:bestfit}
\end{figure*}

Orbital parameters were determined from the detrended light curves using the \jktebop\footnote{http://www.astro.keele.ac.uk/jkt/codes/jktebop.html} orbit-fitting code \citep{southworth2004, southworth2007}. The code is based on the Eclipsing Binary Orbit Program \citep{popper1981, etzel1981}, which relies on the Nelson-Davis-Etzel biaxial ellipsoidal model for well-detached EBs \citep{nd1972, etzel1975}. \jktebop\ models the two components as biaxial spheroids for the calculation of the reflection and ellipsoidal effects, and as spheres for the eclipse shapes. 

Our procedure of removing the out-of-eclipse variability also eliminates gravity darkening, reflected light, and ellipsoidal effects from the light curves. As such, parameters related to these effects are not included in the \jktebop\ modeling. Additionally, out-of-eclipse observations are masked in order to reduce the effect these observations have on the $\chi^2$ calculation and to expedite the fitting process. The RMS in the out of eclipse observations is taken as the constant observational error.

The code finds the best-fit model to a light curve through Levenberg-Marquardt (L-M) optimization. The initial L-M fitting procedure requires reasonable estimates of the orbital parameters to be determined. Period estimates were obtained using Lomb-Scargle \citep{lomb1976, scargle1982} and Box-fitting Least Squares \citep{kovacs2002} periodogram analyses. Approximations of the ephemeris timebase, $T_0$, were obtained by manually phase-folding the light curves on the periodogram period. 

Holding the period and ephemeris timebase fixed, initial L-M fits are performed in succession for the remaining orbital parameters: the surface brightness ratio, $J=(T_\mathrm{eff,2}/T_\mathrm{eff,1})^4$ (which can be approximated by the ratio of the eclipse depths for circular orbits), the sum of the relative radii, $(R_1+R_2)/a$, the ratio of the radii, $k=R_2/R_1$, the orbital inclination, $i$, and the quantities $e\cos\omega$ and $e\sin\omega$, where $e$ and $\omega$ are the eccentricity and periastron longitude, respectively. In systems where contaminating light from neighboring stars is suspected, the so-called ``third light'' parameter, $l_3$, is also investigated as a free parameter. The third light parameter is defined as a constant, such that the sum of the total system light is unity in the out-of-eclipse portions of the light curve. Additionally, in \S~\ref{subsec:star387}, we incorporate radial velocities (RVs) in the fitting procedure, introducing free parameters corresponding to the RV semi-amplitudes of each star in an EB ($K_1$, $K_2$), and the systemic RV, $\gamma$. Analysis of the RVs produces a precise estimate of the mass ratio, $q=M_2/M_1$.

After successively increasing the number of free parameters in the fit, a final L-M fit was performed allowing all relevant parameters to be free. In modeling each system, we assumed a linear limb-darkening law for both components and held the limb-darkening coefficients fixed at reasonable values, discussed further in \S~\ref{sec:individual}. 

The integration times of \Kepler\ long cadence data are comparable to the eclipse durations, resulting in ``phase-smearing'' of the light curve. The long exposure times were accounted for in \jktebop\ by numerically integrating the model light curves at ten points in a total time interval of 1766 seconds, corresponding to the \Kepler\ long cadence duration.

Robust statistical errors on the best-fit model parameters are then found through repeated Monte Carlo (MC) simulations in which Gaussian white noise commensurate to the observational errors is added to the best-fit model. A new L-M fit is performed on the perturbed best-fit model and the new parameters are saved as links in the MC chain. The final orbital parameters for each system are then given by the original L-M best-fit, with uncertainties given by the standard deviations determined from the MC parameter distributions. 

Figure~\ref{fig:bestfit} shows the detrended and phased $K2$ photometry and the best-fit \jktebop\ models, while Tables~\ref{tab:epic387table},~\ref{tab:epic608table}, and~\ref{tab:epic597table} present final values and uncertainties for the fitted orbital parameters derived from corresponding parameter distributions. We note that there are many plausible and excellent fits to the light curves from a statistical robustness perspective, and the L-M approach constrains the parameter combinations based on mutual satisfaction of standard $\chi^2$ constraints.

\section{Overview of System and Primary/Secondary Parameter Estimation}
\label{sec:stellarparams}
\begin{figure*}
\centering
\includegraphics[width=0.33\textwidth]{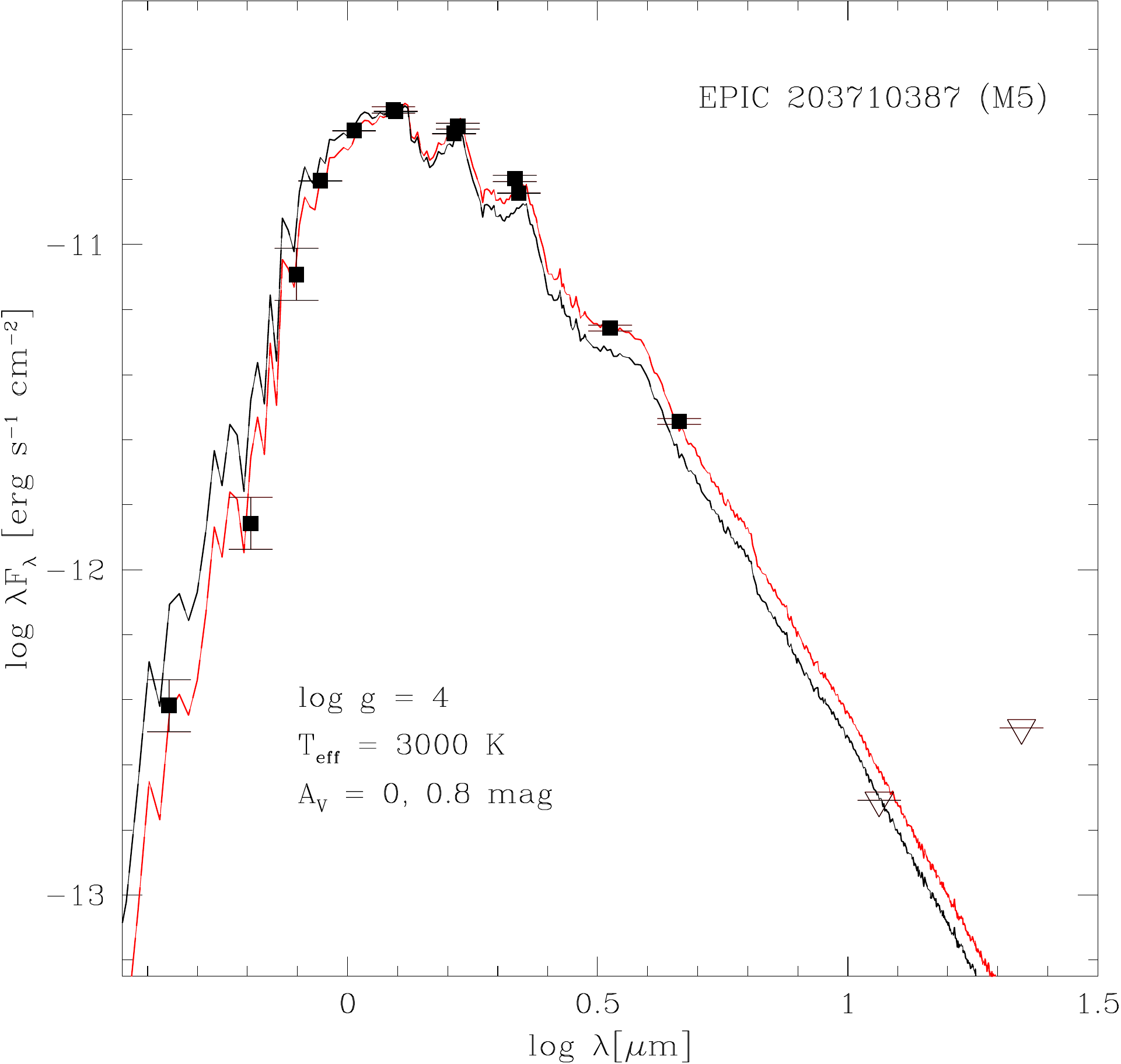}
\includegraphics[width=0.33\textwidth]{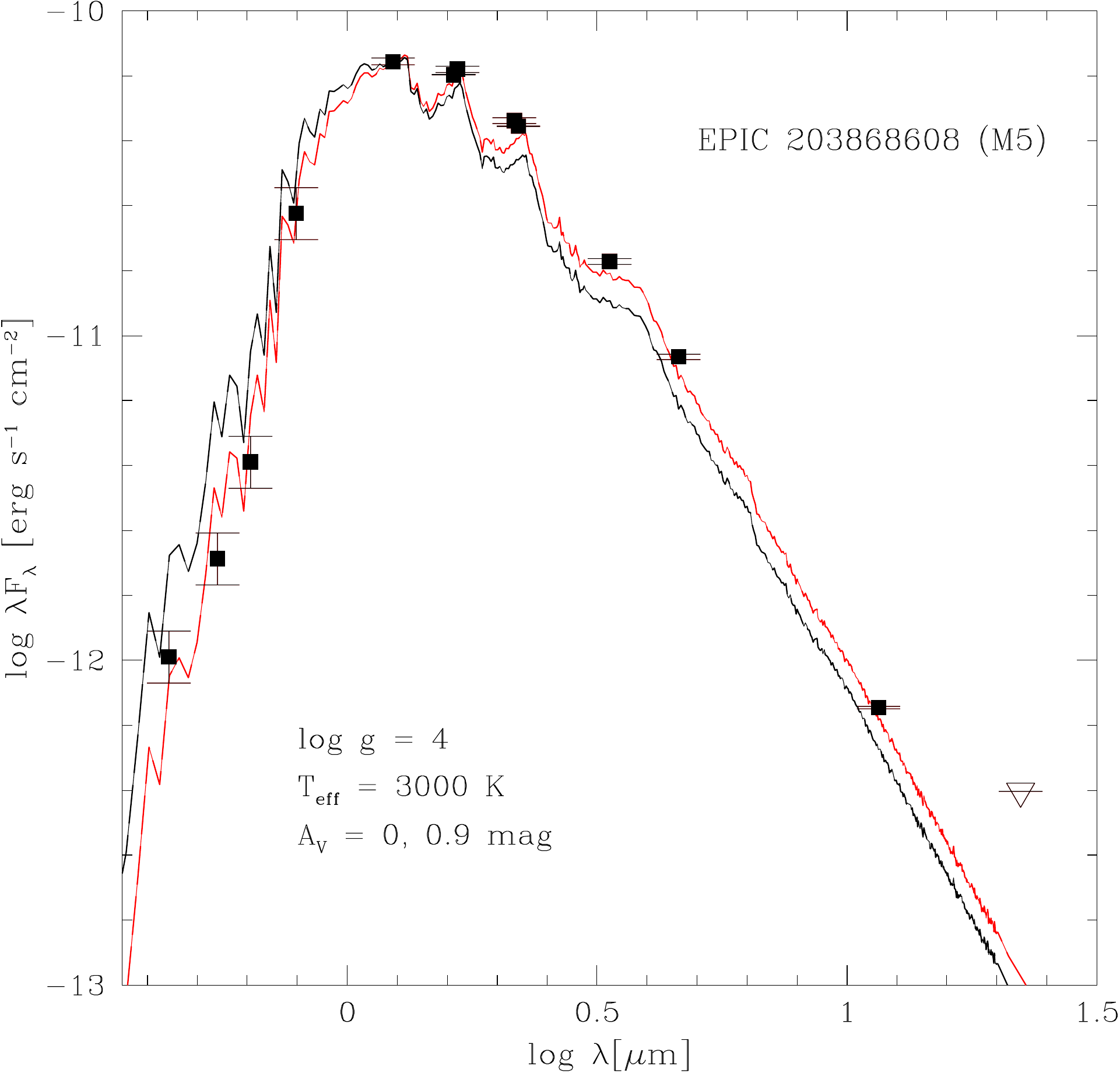}
\includegraphics[width=0.33\textwidth]{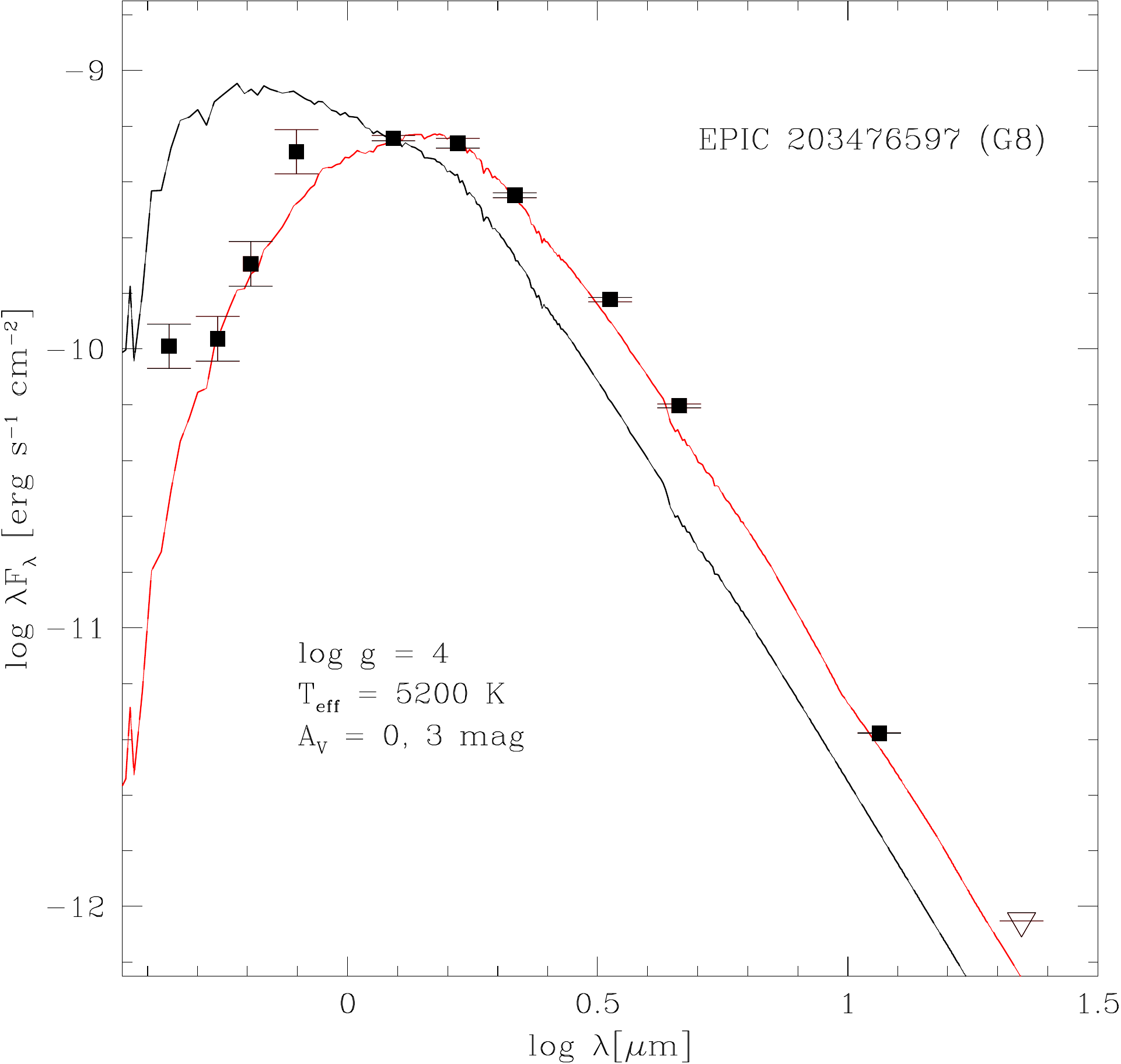}
\caption{Available USNO $BV$, 2MASS $JHK$, UKIDSS $ZYJHK$, and WISE $W1$, $W2$, $W3$, $W4$ photometry or 1$\sigma$ upper limits (downward pointing triangles) compared to NextGen2 model atmospheres. For both \starB\ and \starC, a model atmosphere with $T_\mathrm{eff}$ = 3000 K and $\log{g}$ = 4.0 fits the photometry well. Adopting $A_V=0.9$ mag (red line) produces a better fit to the photometry than an unreddened photosphere (black line). Although the stars have the same spectral type, and \starB\ is a clear double-line system with approximately equal size/temperature and therefore presumably luminosity components, \starC\ is the brighter source. For \starA\ a 5200 K model is adopted, requiring $A_V=3.0$ mag (red line) to match the SED. The $A_V$ values illustrated here are refined in the text based on a match to $J-H$ colors.}
\label{fig:sed}
\end{figure*}

For each of our eclipsing binary systems we have collected available catalog and literature data to assess membership and stellar/disk parameters, as reported in Tables~\ref{tab:epic387table}, ~\ref{tab:epic608table}, and ~\ref{tab:epic597table} and in the discussion below. Spectral energy distributions (SEDs) constructed from broadband photometry for each system are presented in Figure~\ref{fig:sed}. We supplement the literature data with our own spectroscopic observations which allow us to establish or validate spectral type, and confirm membership. 

An important discriminant between likely members and probable non-members in young clusters and moving groups is kinematic information. \cite{lodieu2013} derived a mean proper motion for the previously claimed Upper Sco cluster members  of $\mu_\alpha=-8.6$ mas yr$^{-1}$ and $\mu_\delta=-19.6$ mas yr$^{-1}$, which they noted as a relative value that differs somewhat from the \citet{dezeeuw1999} Hipparcos value on an absolute astrometric frame of $\mu_\alpha =-11$ mas yr$^{-1}$ and $\mu_\delta=-25$ mas yr$^{-1}$. To assess membership likelihood we made use of proper motions reported in the UCAC4 \citep{zacharias2013}  and/or PPMXL \citep{roeser2010} catalogs. 
We further assess membership based on radial velocities from the HIRES data.
The details for the individual EB systems are discussed below.

Finally, for all three EB systems, the H$\alpha$ emission and \ion{Li}{1} 6707.8 \AA\ absorption line strengths illustrated in Figure~\ref{fig:spectra}, discussed above, are consistent with the expectations for young active low mass stars.  

Having assessed membership, we used literature and our own HIRES-derived spectral types to estimate effective temperature based on empirical calibrations for pre-MS stars, and then incorporated 2MASS photometry to calculate combined system luminosities.  The near-infrared colors of all three sources are slightly redder than expected from young star intrinsic colors, suggesting a modest amount ($A_V\sim$ 1-3 mag) of reddening.  From the spectral type and broadband SED we calculated the extinction, and then the corresponding $J$-band based luminosity (which also assumes the cluster distance, here assumed to be the \cite{dezeeuw1999} value of 145$\pm$13 pc). 
For those systems in which we could measure the radial velocities of both eclipsing components (EPIC 203710387 and EPIC 203868608), we directly determined the masses and radii through mutual fitting of the light curves, radial velocity time series, and spectroscopic flux ratios. In these cases, distance-independent luminosities are determined from the temperatures (based on spectral types) and radii using the Stefan-Boltzmann law and an assumed value of $T_\odot = 5771.8 \pm 0.7$ K\footnote{From the total solar irradiance \citep{kopp2011}, the solar radius \citep{haberreiter2008}, the IAU 2009 definition of the AU, and the CODATA 2010 value of the Stefan-Boltzmann constant.}. For EPIC 203476597, in which the secondary lines were too weak to measure reliable radial velocities, we estimated the primary radius from its projected rotational velocity, $v\sin{i}$, and rotational period. The secondary radius is then determined from the orbit-fitting results. Model-dependent masses for the components of this system are derived from interpolation between PARSEC models \citep{bressan2012}. We used either PARSEC or \cite{baraffe2015}, hereafter BHAC15, pre-main sequence evolutionary models to also estimate the ages of each system.

\section{Results and Discussion of Individual Eclipsing Binaries}
\label{sec:individual}

For each system we now discuss characteristics of the raw $K2$ photometry, the details of the light curve detrending procedure, and the results of the orbital and stellar parameter determinations using methods described above. In each case, we have sampled a range of possible orbital parameter fits and assessed Monte Carlo parameter distributions of all fitted parameters, in some cases needing to constrain or fix certain parameters in order to produce physically reasonable overall solutions. For each fitted parameter, uncertainties are derived from Monte Carlo error propagation after including the uncertainties in anchoring stellar properties.

\subsection{EPIC 203710387}
\label{subsec:star387}

This system is comprised of nearly identical M4.5 or M5 components in a circular orbit with period $\approx$2.8 d. Separated by $\sim$5 \rsun, or $\sim$11 stellar radii, and inclined $\sim$83$^\circ$ to our line-of-sight, the stars undergo partial or grazing eclipses of nearly equal depths. As noted earlier, the system is double-lined, and radial velocity measurements indicate approximately equal mass components.  

A 1.5-pixel radius aperture was found to produce the highest quality $K2$ light curve. Primary and secondary eclipses of $\sim$7\% depth each are notable in the raw photometry (see Fig.~\ref{fig:lightcurve}).  The low-pass flux is a significant improvement over the raw flux, but only in the first half of the campaign where the centroid drifts are smaller. Over the entire campaign, the low-pass flux has a higher noise level than the raw photometry, so in this instance the raw photometry was selected for further correction. 

After applying the detrending procedure, the phased light curve was divided into 100 bins. In each phase bin the mean flux and standard deviation were computed and 3-$\sigma$ outliers were identified, resulting in the exclusion of an additional 38 observations across the entire campaign.

The 2.5 hr and 6.5 hr quasi-CDPP of the detrended light curve (including in-transit observations) are 1313 ppm and 886 ppm, respectively. In this case, the detrending provides a $\sim$20\% improvement over the raw photometry on 6.5 hr timescales or $\sim$5\% improvement on 2.5 hr timescales.  Inspection of the broader dataset revealed that hundreds of other light curves from Campaign 2 display the same variation in quality between the first and second halves of the time series. Though considering only observations from the first half yields a more precise light curve, we opted to include all of the observations for the benefit of sampling additional transits.

The star is included in Table 1 of \cite{lm2012}, which lists properties of known Upper Sco members, however, there is a lack of literature on this source prior to that work. The star was then identified as a candidate member by \cite{lodieu2013} based on both its proper motion and location in an infrared color-magnitude diagram. The two independent measurements of the proper motion, ($\mu_\alpha, \mu_\delta = -11.8 \pm 5.1, -28.0 \pm 5.1$ mas~yr$^{-1}$) from \cite{roeser2010} and ($\mu_\alpha, \mu_\delta = -12.30 \pm 1.82, -19.96 \pm 1.82$ mas~yr$^{-1}$) from \cite{lodieu2013},\footnote{Notably, proper motion measurements for \starB\ are not included in UCAC4 \citep{zacharias2013} or URAT1 \citep{zacharias2015}.} are consistent with one another, and with the mean values among Upper Sco members with $\chi^2 < 1-2$ (depending on which values are adopted).  There is no evidence for circumstellar material around \starB, with the object too faint for WISE in its two longest bands.  The location is south and west of the main $\rho$ Oph cluster, in a relatively lower extinction region.

An M5 spectral type  was reported by \cite{lm2012}. From our HIRES spectrum we estimate a spectral type of M4.5 and report both H$\alpha$ emission and lithium absorption, confirming the youth of \starB\ (see Fig.~\ref{fig:spectra}).  As an external consistency check, we constructed an SED from the available broadband photometry and compared it with artificially reddened NextGen2 model atmospheres based on \cite{hauschildt1999} to find plausible combinations of spectral type and $A_V$ (see Fig.~\ref{fig:sed}). We found that a model atmosphere having \teff\ = 3000 K (corresponding to approximately spectral type M5), $\log{g}$ = 4.0, and $A_V$ = 0.8 mag provides a good match to the broadband photometry, though we refine both the temperature and extinction below. 

We also compared broadband colors with the empirical spectral type - color - temperature relations of HH15 and \cite{pm2013}, hereafter PM13. The $J-H$ color evolves rapidly in the pre-main sequence and is not well-reproduced by evolutionary models. The 2MASS $J-H=0.655\pm0.033$ mag color of \starB\ is consistent on the HH15 color scale with an M4 spectral type if similar to `young' 3-8 Myr old moving group members, but an M0-M3 spectral type if more similar to `old' 20-30 Myr moving group members. On the PM13 color scale appropriate for 5-30 Myr old stars, the $J-H$ color suggests an M2-M4 star.   The quoted UKIDSS photometry produces $J-H=0.567\pm0.002$, more consistent with a young M5 star according to both HH15 and PM13.  Allowance for a small amount of reddening would argue for earlier spectral types on these color scales.

We conclude that the near-infrared photometry as well as the broader SED are consistent with the previously determined M5 spectral type for a young pre-main sequence age, so we adopt this spectral type in what follows. The corresponding effective temperature from HH15 (their Table 2) is \teff\ = 2980 K, or from PM13 \teff\ = 2880 K.  We adopt the former along with an uncertainty of $\pm$ 75 K in \teff\ to account for a possible 0.5 subclass error in the spectral type, as suggested by \cite{hh2015}. 

We find from the photometry that $A_V=1.2\pm0.3$ mag and the $J$-band based system luminosity $\log{(L/L_\odot)}=-1.64 \pm 0.08$, where the error terms come from Monte Carlo sampling of the allowed error in temperature, but for luminosity are dominated by the uncertainty in the distance. This calculation places the object in the middle of the Upper Sco temperature-luminosity sequence, reaffirming its presumed youth and membership.  From the luminosity estimates and the plausible effective temperature range, and assuming equal luminosity components (consistent with the nearly equal mass components), we estimate the individual stellar radii at $0.40 \pm 0.04$ \rsun\ each. Direct radii measurements are derived later from combination of the light curve and RVs, but are broadly consistent with this approximation. 

\begin{figure*}
\centering
\includegraphics[width=0.99\textwidth]{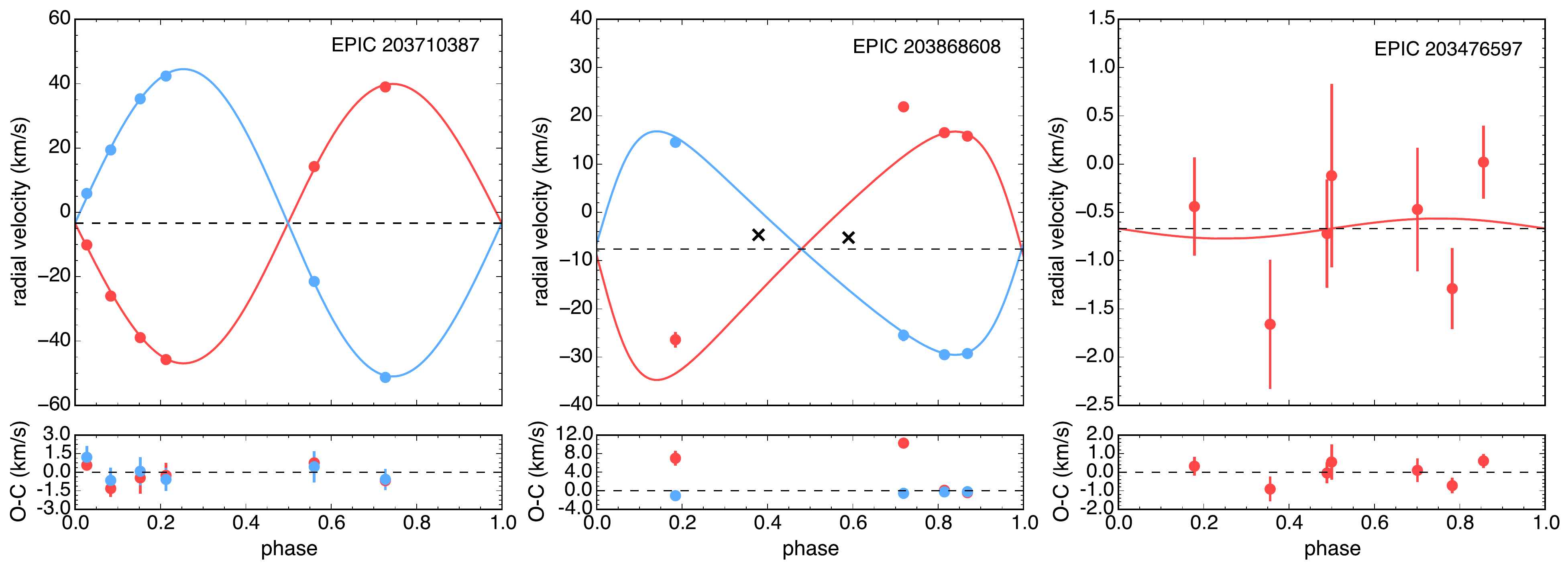}
\caption{For each EB studied here, the radial velocity curve (upper panel) and best-fit residuals (lower panel). The measurements are phase-folded on the best-fit period from simultaneous fitting of RVs and the $K2$ light curve with \jktebop. The red and blue points and curves are the observations and best-fit model, for the primary and secondary components, respectively. Each point indicates the weighted mean radial velocity derived from measurements over several spectral orders within a single spectrum. Each measurement receives a weight equal to the inverse of the variance. The error bars represent the corresponding standard deviation between the multiple measurements, which in the top panel are smaller than the points themselves. In the case of EPIC 203868608, which is a triple system, two measurements at essentially the mean systemic velocity of Upper Sco ($\sim$-4 \kms) are indicated by the black crosses. These measurements are likely compromised due to the low expected velocity separation that is comparable to the spectrograph resolution, and were consequently excluded from the RV fits in order to obtain a good fit.}
\label{fig:rv}
\end{figure*}

The M4.5 radial velocity standard GJ388 \citep{nidever2002}, was used to measure absolute RVs from the HIRES spectrum. Several spectral orders with high signal-to-noise were chosen to produce multiple measurements per observation. In the orbital parameter fitting, an individual RV measurement for each epoch was derived from a weighted average of individual measurements from separate orders of the spectrograph. The \jktebop\ code is capable of fitting light curves and RV curves simultaneously, but only considering one RV curve at a time. The systemic velocities for each component were forced to be equal, and the resulting best-fit value ($\gamma \sim$ -3 \kms) is consistent with values typical of Upper Sco members \citep{dezeeuw1999, mohanty2004, kurosawa2006}. The radial velocity curves for both components and best-fitting models are shown in Figure~\ref{fig:rv}.

It is typically considered good practice to allow limb darkening coefficients to be free parameters when fitting light curves, given that these coefficients are largely uncalibrated \citep{southworth2007}. However, as noted by \cite{gillen2014}, \jktebop\ is susceptible to allowing non-physical limb-darkening parameters to find a good fit. Furthermore, grazing eclipses do not contain enough information to constrain the limb-darkening coefficients. We find that allowing the limb-darkening parameters to vary does not change the other fitted parameters significantly, so we hold the limb-darkening coefficients fixed. We assumed a linear limb-darkening law for both components, setting the coefficient $u=0.888$, corresponding to the mean of all values calculated by \cite{claret2012} satisfying 2780 K $\leq T_\mathrm{eff} \leq$ 3180 K and 4.0 dex $\leq \log{g} \leq$ 4.5 dex, appropriate for an M4.5-M5 PMS star. 

Initial attempts to fit the photometry to model light curves revealed strong degeneracies between the stellar radii related parameters, inclination, and surface brightness ratio, in large part due to the quite poorly constrained parameter, $k=R_2/R_1$. This is expected: for detached EBs with similar components in a grazing configuration, the sum of the fractional radii is well-defined (depending mainly on the inclination and eclipse durations), but the eclipse shapes are relatively insensitive to the ratio of the radii \citep{andersen1980, southworth2007}.

The degeneracy is so strong that allowing $R_2/R_1$ to be a free parameter resulted in a best-fit that suggested nearly equal mass components ($<10\%$ difference in the masses, which are well constrained by the RVs) but with a $\sim$35\% difference in the radii, such that the more massive component was smaller.  Although this solution provided a good fit, it implied a physically unlikely scenario in which the more massive component would be nearly a factor of 10 older than the secondary when compared to BHAC15 mass-radius isochrones.  Notably, however, the large uncertainties did admit that the radii were consistent within $2\sigma$ with being equal size. 

\begin{figure*}
\centering
\includegraphics[width=0.99\textwidth]{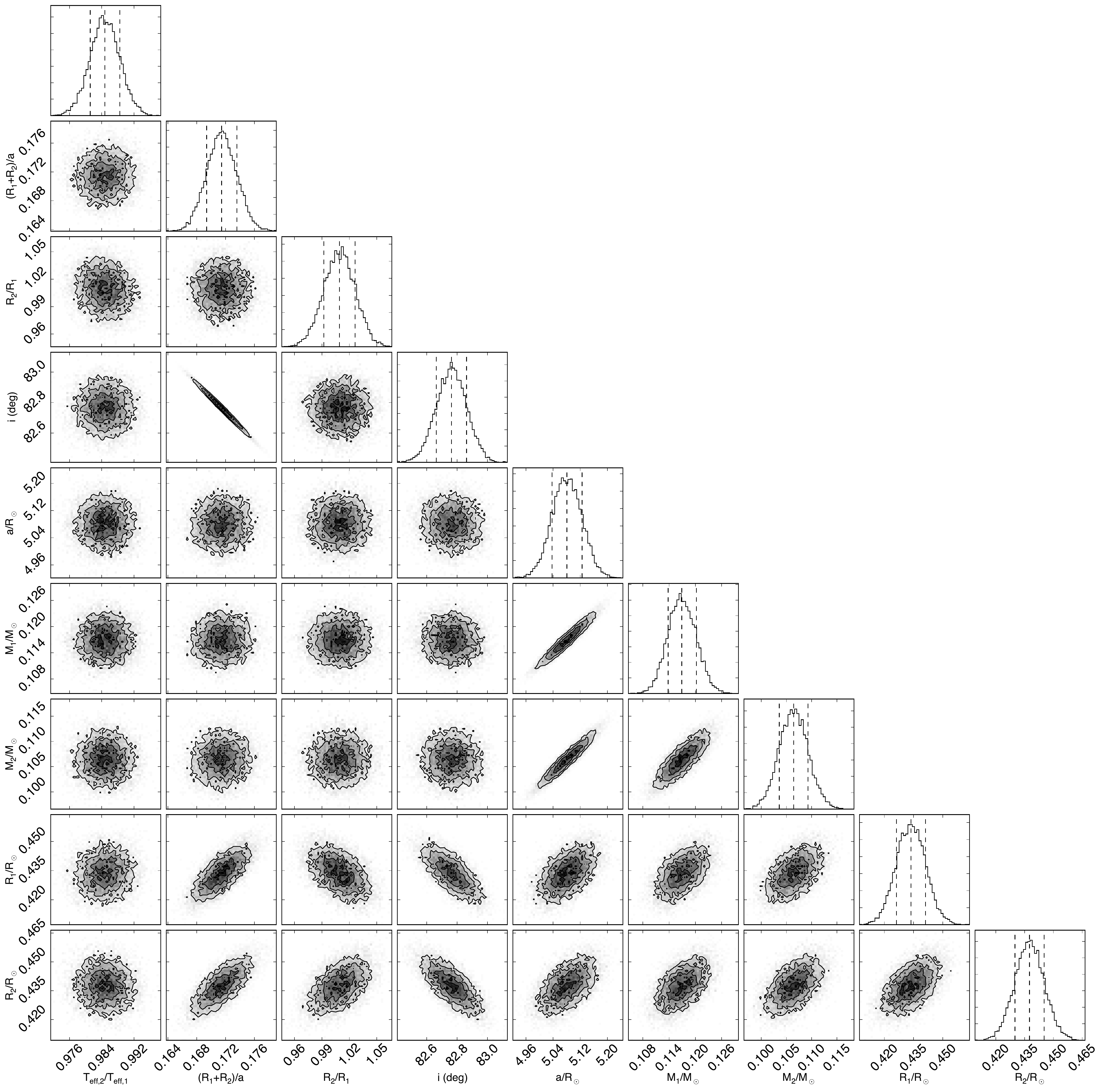}
\caption{ Distributions of selected free and derived parameters and their pairs from the MC fitting procedure in the circular orbit fit for \starB. The 0.5-, 1.0-, 1.5-, and 2.0-$\sigma$ contours are drawn. The dashed lines in the 1D parameter distributions represent the median and 68\% confidence intervals of the distribution. This plot was created using the \texttt{triangle} {\sc Python} code (Foreman-Mackey et al. 2014, DOI:10.5281/zenodo.11020).}
\label{fig:staircase}
\end{figure*}

From the HIRES spectra, we measured spectroscopic flux ratios at each epoch from the relative heights of the two distinct cross-correlation function peaks (presented in Table~\ref{table:rvs}). We provide these flux ratio time series data as input in the final modeling with \jktebop\ which effectively breaks the degeneracy in the ratio of radii noted above. The final orbital parameters (including masses and radii, given the presence of RVs), which are the result of 5,000 MC simulations with \jktebop\ are presented in Table~\ref{tab:epic387table}. Figure~\ref{fig:staircase} shows distributions of selected parameters derived from the MC fitting procedure.  

We explored a solution to the light curve and radial velocities in which $e\cos\omega$ and $e\sin\omega$ were adjusted parameters, as well as one in which the eccentricity is assumed zero. These two solutions (presented in Table~\ref{tab:epic387table}) show very good agreement in most adjusted and derived parameters, and indeed the best-fit eccentricity is within 2-$\sigma$ of zero. However, the $\chi^2_\mathrm{red}$ is significantly lower in the eccentric case, and close inspection of light curve (see Fig.~\ref{fig:lightcurve}) confirms that the secondary eclipse occurs just slightly before phase=0.5. As such, we ultimately adopt the eccentric orbital parameters and subsequently derived quantities in our final analysis. 
Though, it is interesting to note that the circular orbit solution leads to a temperature ratio much closer to unity, $T_\mathrm{eff,2}/T_\mathrm{eff,1}=0.984\pm0.004$, relative to the temperature ratio favored by the eccentric solution, $T_\mathrm{eff,2}/T_\mathrm{eff,1}=0.953\pm0.019$. Both solutions suggest the secondary is larger than the primary, though the circular solution favors a ratio of radii very close to one, $k=1.009\pm0.017$, compared with the eccentric solution value of $k=1.077\pm0.045$. Notably, with such a short orbital period and well-constrained age, this system should be quite valuable for studies of pre-MS circularization timescales.

The final masses and radii of the two components of \starB, resulting from the equal-radii light curve solution, suggest an age of $\sim$ 10-11 Myr for the system when adopting the BHAC15 models and circular solution parameters (the median ages and 1-$\sigma$ errors are $11.6\pm0.4$ Myr for the primary, and $9.9\pm0.3$ Myr for the secondary, from interpolation between isochrones in the mass-radius plane). If we consider a more traditional age for Upper Sco of 5 Myr, the implication is that the BHAC15 models over-predict the radii by $\sim 25-35\%$ for a given mass and age.

From the bolometric luminosity, the best-fit luminosity ratio ($L_2/L_1 \approx k^2 J$, for circular orbits), and the directly determined stellar radii we can compute the effective temperatures of each component. We calculated $T_\mathrm{eff,1}=2940\pm150$ K and $T_\mathrm{eff,2}=2800\pm150$ K, where the uncertainties come from standard error propagation. The placement of each component in \teff-$\log{g}$ space relative to BHAC15 isochrones is consistent with an age of $\sim$11-14 Myr, though the corresponding model masses are underestimated by a factor of 2. Allowing for temperatures $\sim$175-200 K hotter, while holding $\log{g}$ fixed, brings the model-predicted masses into better agreement with the dynamical measurements and lowers the age of each component by $\sim$1 Myr. If we instead assume a primary temperature from the spectral type, we obtain temperatures of $T_\mathrm{eff,1}=2980\pm75$ K and $T_\mathrm{eff,2}=2840\pm90$ K, which helps to resolve some of the model discrepancies in mass and age noted above. Assuming these temperatures and the directly measured radii, we then calculate \emph{distance-independent} luminosities of $L_1=0.0124\pm0.0014$ \lsun\ and $L_2=0.0119\pm0.0016$ \lsun. We ultimately adopt the temperatures based on the spectral type and the distance-independent luminosities in our final analysis.

With a period of approximately 2.8 d and a separation of only $\sim$11 stellar radii, the system is quite compact. However, it still meets the criterion for detachment. Using the precise mass ratio derived from RVs, we calculate the effective Roche lobe radius for the system to be $\approx$ 37\% of the separation, or $\approx$1.9 \rsun, from the formula of \cite{eggleton1983}.

Eclipsing binary light curves, and thus the parameters derived from them, are susceptible to the level of extraneous light from other stars in the photometric aperture. This contamination from nearby sources, whether associated or not, is known as third light in the EB literature. The effect of third light on EB light curves is to decrease the depths of eclipses and mimic a system with lower inclination \citep{kallrath2009}. To assess potential sources of contamination, \starB\ was imaged with Keck/NIRC2 in a $K_p$ (2.12 $\mu$m) filter on May 27, 2015 UT. The dithered mosaic covered a $15''\times15''$ region, but due to the dither pattern used, the upper $\sim$ 5.8\arcsec$\times$5.8\arcsec region in the northeast quadrant of the mosaic was not covered. A star was detected that is 3.6 mag fainter than \starB, at a position angle of $\sim$332\arcdeg\ measured east of north and separation of 1.6$''$. This nearby source is unaccounted for in the light curve modeling, but likely contaminates the $K2$ photometry at the few percent level. 

We explored the possibility of fixing the third light parameter at 3.6\%, corresponding to the contamination in $K_p$. This trial resulted in a slightly higher $\chi^2_\mathrm{red}$ than our best-fit solution presented in Table~\ref{tab:epic387table}, and masses and radii that change within error of our reported values. As such, we choose to ignore third light for this system but note it may indeed introduce an additional few percent uncertainty in the absolute radii, though not nearly enough to favor an age as young as 5 Myr in the mass-radius plane.

\begin{figure}
\centering
\includegraphics[width=0.475\textwidth]{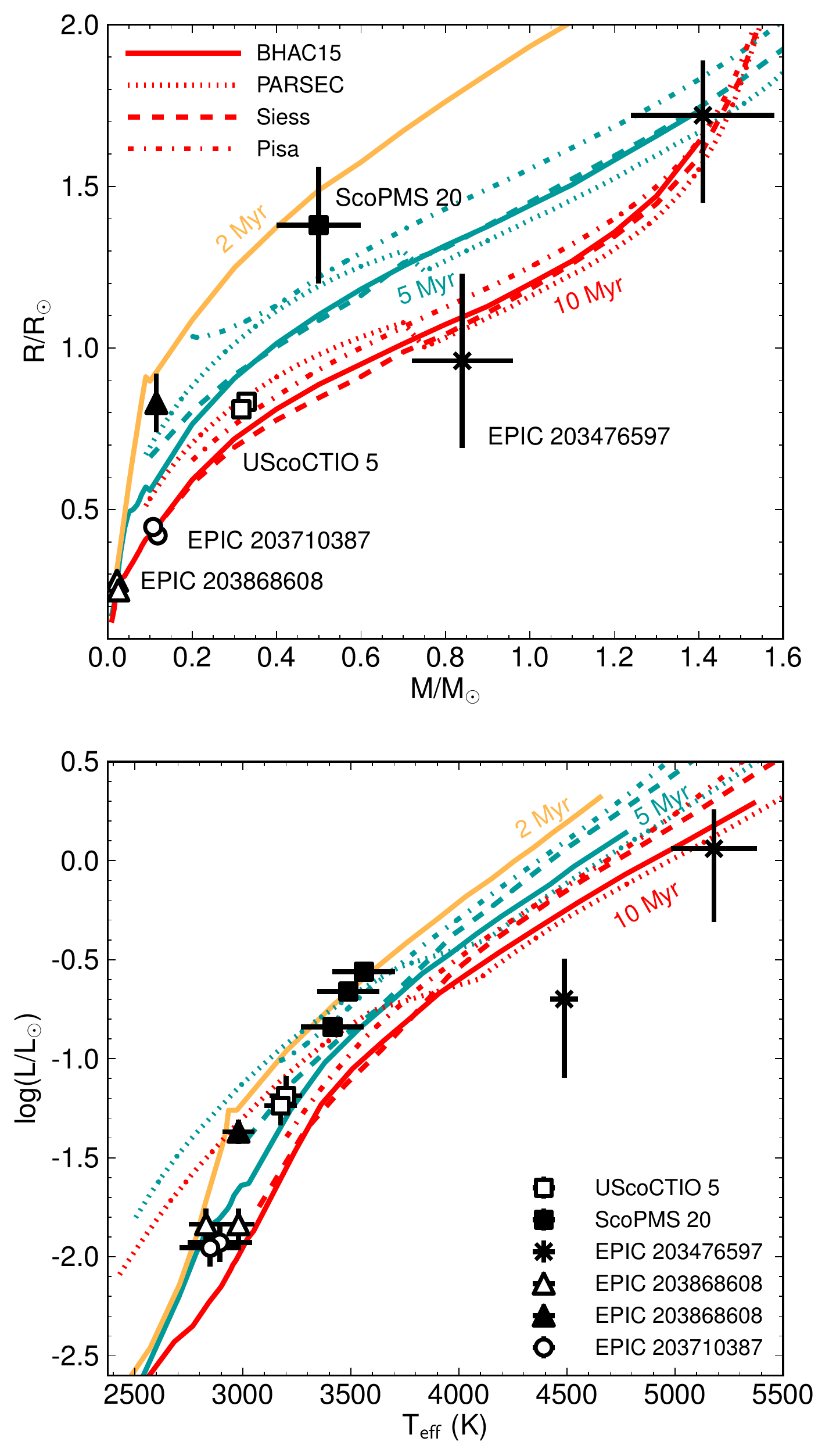}
\caption{Isochrones in the mass-radius (top) and temperature-luminosity (bottom) planes with the three EBs discussed here and two other low-mass systems in Upper Sco: both components of UScoCTIO 5 \citep{kraus2015} and the primary of the triple system ScoPMS 20 \citep{mace2012}. The BHAC15, PARSEC v1.2s \citep{bressan2012, chen2014}, \cite{siess2000}, and Pisa \citep{tognelli2011} pre-MS evolutionary models at solar metallicity (Z=0.02) are considered for comparison. The two components of \starB\ are overlapping in the mass-radius plane. UScoCTIO 5 and \starB\ have fundamentally determined masses and radii; errors are smaller than the points themselves. The eclipsing components of \starC\ also have fundamentally determined masses and radii, though large uncertainties remain for this system, particularly in the luminosities, for the reasons discussed in \S~\ref{subsec:star608}. The tertiary of this system does not have fundamentally determined parameters and hence is represented by the filled black triangle. All other systems have parameters that depend on models and/or empirical relations. In the lower panel, the equal-temperature, equal-luminosity components of UScoCTIO 5 are offset for clarity. No single isochrone can reproduce the fundamentally determined masses and radii of both the \starB\ and the UScoCTIO 5 systems.}
\label{fig:isochrones}
\end{figure}

\subsection{EPIC 203868608}
\label{subsec:star608}

High-angular resolution imaging revealed that this system is likely a hierarchical triple, with the EB components in an eccentric 4.5 d orbit and an M-type companion within 20 AU. Orbital motion of tens of \kms\ was detected with six epochs of Keck-I/HIRES spectroscopy, indicating the M4.5-M5 type which dominates the spectrum must be the primary component of the EB. The system is double-lined, and though we find a good model fit for only one of the RV curves (see Fig.~\ref{fig:rv}), there is compelling evidence that the sum of the RV semi-amplitudes is $<$60 \kms, corresponding to a total system mass of $M_1+M_2 \lesssim$ 0.1 \msun. If confirmed, this would constitute only the second double-lined eclipsing brown dwarf binary to date, the first being 2MASS J05352184–0546085 in Orion \citep{stassun2006, stassun2007}.

A 2-pixel aperture produced the highest quality $K2$ photometry, and the raw flux (rather than the low-pass flux) was selected for further correction. The light curve exhibits both narrower 12.5\% primary and broader 10\% secondary eclipses, indicating a non-negligible eccentricity, with period 4.54 days.  There is also a superposed sinusoidal pattern likely due to rotation with a period just over 1 day, as well as longer time scale variations. 

After applying the detrending procedure, the phased light curve was divided into 100 bins. In each phase bin the mean flux and standard deviation were computed and 3-$\sigma$ outliers were identified, resulting in the exclusion of an additional 54 observations across the entire campaign. The observational errors were determined from the RMS scatter of the out-of-eclipse observations taken during the first half of the campaign, which have a slightly higher noise level than the second half for this particular system.

The colors of the primary are quite red, corresponding to a late M spectral type, with an M5 star consistent with the HIRES spectrum.  Proper motion is not available in UCAC4 but the values in PPMXL ($\mu_\alpha, \mu_\delta = 1.3, -19.6$ mas~yr$^{-1}$) are inconsistent with Sco membership at the $>3.5\sigma$ level, perhaps due to astrometric contamination from the faint closely projected companion. Nevertheless, our detection of Li absorption and H$\alpha$ emission confirm the youth of the system. The source is located due west of the embedded $\rho$ Oph cluster, close to \starB\ in fact.

As with \starB, we adopt an effective temperature of \teff=2980$\pm$75 K, from the empirical calibration of HH15. Monte Carlo error propagation of 50,000 points drawn from a normal distribution in \teff\ was used to determine $A_V$ and bolometric luminosity. 

We adopt the same analysis approach used for \starB\ and assume a linear limb-darkening law for both EB components, setting the coefficient $u=0.888$, corresponding to the mean of all values calculated by \cite{claret2012} satisfying 2780 K $\leq T_\mathrm{eff} \leq$ 3180 K and 4.0 dex $\leq \log{g} \leq$ 4.5 dex, appropriate for an M4.5-M5 PMS star.

Radial velocities were acquired over six epochs with HIRES. Though the system is double-lined, only one velocity component (at roughly -5 \kms) could be extracted from the spectra at two epochs, corresponding to phases $\sim$0.4 and 0.6. Somewhat conspicuously, these two epochs are approximately equidistant in phase from the predicted time of secondary eclipse, as demonstrated in Figure~\ref{fig:rv}. The expected velocity separation at these epochs is only a few times the resolution of the spectrograph and we were unable to distinguish two peaks in the cross-correlation function, only a single peak with the quoted velocity, which is near systemic for the binary. In our final mutual fit of the $K2$ light curve and HIRES RVs, we exclude these two discrepant observations, which are not obviously associated with either component. We also measured spectroscopic flux ratios from the HIRES data for each of the four epochs included in the radial velocity fitting. We estimated the flux ratios from the relative heights of the two distinct cross-correlation function peaks. These flux ratios were included as input in the \jktebop\ modeling and helped to constrain the ratio of radii.

The total system luminosity is $\log{(L_\mathrm{bol}/L_\odot)}=-1.14 \pm 0.08$, where the uncertainty is dominated by the uncertainty in distance.  Despite their similar spectral types, \starC\ has a luminosity that is larger by a factor of $\sim$3 than that of \starB. The cluster distance was assumed in the luminosity calculations, and given the significant cluster depth ($\sigma_d/d \approx$ 9\%), we considered the possibility that different distances could account for some of the luminosity discrepancy. If we allow for a 2-$\sigma_d$ separation in the line-of-sight distance between the two systems, \starC\ is still more than twice as luminous as \starB. We therefore conclude that differing distances is unlikely to account for the entire luminosity discrepancy between the two systems.

High angular resolution imaging of the system revealed nearby sources which partially resolves the luminosity discrepancy noted above. A snapshot image taken on July 14, 2015 UT with the MAGIQ guide camera on Keck/HIRES revealed a fainter source with $\Delta m$ = 3.58 $\pm$ 0.10 mag that at $\sim$4\arcsec\  separation is blended with \starC\ in the $K2$ aperture. Additionally, there is a source $<15$\arcsec\ to the southeast with $\Delta m$ = 1.45 $\pm$ 0.01 mag fainter than \starC\ that is partially enclosed by the $K2$ aperture (see the middle right panel of Fig.~\ref{fig:stamp}, in which this source is resolved in DSS2).  

Keck/NIRC2 images obtained on July 25, 2015 UT then revealed a nearly equal brightness ($\Delta J = 0.278 \pm 0.034$ mag, $\Delta K_p = 0.316 \pm 0.021$ mag) companion at a projected separation of 0.12\arcsec. At the distance of Upper Sco, this corresponds to a separation of $<20$ AU, indicating the second source is likely a bound companion. We then propose that the eclipses provide evidence that \starC\ is a hierarchical triple system. Additionally, there exists a more widely separated source to the southeast with $\Delta K_p = 5.11 \pm 0.021$ mag fainter than the brighter component of the nearly equal-brightness pair.

For the remaining discussion, we will assume a primary is being eclipsed by a secondary to remain consistent with the language used to this point. We then designate the third more distant, and presumably single, companion as the tertiary. 
The NIRC2 imaging indicates a flux (and thus luminosity) ratio between the two near-equal-brightness components of $\sim$0.75-0.80 in $J$-band or $\sim$0.73-0.76 in $K_p$. For the remaining discussion, we will approximate the NIRC2 flux ratio as 0.765$\pm$0.035, the mean of the lower limit set by the $K_p$ band and the $J$-band upper limit. At present, however, we can not say with certainty which component of the pair is the presumably single star and which is the EB. Thus, we consider two general scenarios: (1) in which the tertiary is more luminous than the combined luminosities of the EB components, and (2) in which the tertiary is fainter than the combined EB luminosity. 

For each of the two scenarios above, we calculate the expected third light parameters given the measured flux ratios of the tertiary and all other contaminating sources within the aperture. We consider the contributions from all of the blended sources discussed above, as well as the relatively bright source that is only partially enclosed by the $K2$ aperture. For this partially enclosed source, we consider a range of values for the fraction of light enclosed by the aperture. In the absence of other data, we assume the contamination from the tertiary in the \Kepler\ bandpass is equal to the measured NIRC2 contamination.

In the first scenario, in which the tertiary luminosity exceeds the EB combined luminosity, the EB contributes approximately 40\% of the total system light in the $K2$ light curve, accounting for both the tertiary and fainter contaminating sources in the aperture. This implies a third light parameter of $l_3=0.59$, if only 5\% of the light from the partially enclosed source is contaminating the aperture, or as high as $l_3=0.63$ if half of the light from this source is enclosed.

When allowing the third light parameter to be free, the light curve fitting favors the first scenario, settling on a best fit with a third light parameter of $l_3=0.684\pm0.016$. Assuming $L_\mathrm{bol}=L_1+L_2+L_3$ (with no additional sources of contamination), and using the NIRC2 measured flux ratio, and EB luminosity ratio, $L_2/L_1$, that arise from the best fit model, we determine EB component luminosities of $L_1=0.0146\pm0.0029$ \lsun, $L_2=0.0146\pm0.0029$, and a tertiary luminosity of $L_3=0.0428\pm0.0060$ \lsun. From combination of the light curve and radial velocities, this fit results in EB component masses and radii of $M_1 = 0.02216 \pm 0.00045$ \msun, $M_2 = 0.02462 \pm 0.00055$ \msun, $R_1 = 0.2823 \pm 0.0051$ \rsun, $R_2 = 0.2551 \pm 0.0036$ \rsun. We note that the luminosities calculated above are overluminous by a factor of three given the measured effective temperatures and radii. For the radii favored by the fit, and the temperatures we measure from the spectral type and $J$, the implied luminosities are $L_1=0.0046\pm0.0005$ \lsun\ and identically $L_2=0.0046\pm0.0005$ \lsun.

However, this scenario, in which the tertiary is more than twice as luminous as either the primary or secondary, is incongruous with the detection of orbital motion of tens of \kms\ in the Keck-I/HIRES spectra.  If the tertiary was so luminous, contributing more than half the total system light, it should be readily detectable as a distinct component from the M4.5-M5 primary which exhibits the large radial velocity shifts.

Nevertheless, for completeness, we calculate the EB parameters implied by the best-fit light curve model in this first scenario. Notably, the best-fit model suggests a ``secondary'' that is slightly hotter, more massive, but smaller than the primary. However, we stress that there are unquantifiable uncertainties due to the fact that the RVs for only one component are well fit by the models.  If we assume the HH15 M5 temperature of $T_\mathrm{eff,1}=2980\pm75$ K for the primary in this scenario, then the radius implied by the luminosity is $R_1\approx0.40\pm 0.04$ \rsun, consistent with the radii of the components of \starB, but discrepant at the 3-$\sigma$ level with the radius implied by our light curve and radial velocity fit.  We note that the spectral type we find is earlier than the M6.5 spectral type of the eclipsing brown dwarf binary found in the younger Orion Nebula \citep{stassun2006, stassun2007}. For comparison, the primary of that system has a mass of $M$=0.054$\pm$0.005 \msun\ and temperature of \teff=2650$\pm$100 K (from the spectral type).

Meanwhile, the HIRES spectrum shows no evidence for a component earlier than M4.5. Thus, the tertiary must have a similar temperature and, given its large luminosity, a radius of $R_3\approx0.83\pm0.09$ \rsun. However, we again emphasize that if the tertiary is indeed so luminous it should have been detected as a distinct peak in the cross-correlation functions.

Evidence in favor of this first scenario is found when comparing the near-IR brightnesses predicted by models for brown dwarfs and stellar mass M-types, with the measured NIRC2 magnitude differences. If we assume the system is composed of a single M5 star (the wide tertiary at $\sim$ 20 AU) with an eclipsing pair that are equal in brightness to each other in either $K$- or $J$-band, we can use the $K$- and $J$-band magnitude differences from NIRC2 to interpolate between evolutionary models and estimate the masses of the eclipsing pair. For example, BHAC15 models predict an M5 star (here approximated as a 0.1 \msun\ star) at 10 Myr should have $K$=6.60 mag, $J$=7.44 mag. Holding the age fixed, two brown dwarfs of $\sim$0.03 \msun\ could reproduce the NIRC2 magnitude differences in either $J$ or $K$. Allowing the age to be as young as 3 Myr would imply an eclipsing pair of $\sim$0.04 \msun\ brown dwarfs.

In the second scenario, the EB combined luminosity is greater than the tertiary luminosity. In this case, the EB contributes approximately 50\% of the total system light in the $K2$ light curve, depending on the fraction of light included from the partially enclosed source. The range of third light parameters corresponding to 5-50\% containment of the partially enclosed source is $l_3=0.46-0.52$.

In this scenario, we can no longer rely on a light curve model that has a third light parameter $>55\%$. We perform a new light curve fit using 1,000 MC simulations and fixing third light to $l_3=0.50$. Possibly supporting this scenario are the NIRC2 $J-K_p$ colors of the components in the equal-brightness pair, which are 1.064 $\pm$ 0.033 and 1.073 $\pm$ 0.023 (both uncorrected for reddening). Such similar colors suggest the tertiary and the EB primary have quite similar temperatures. However, we note that the BHAC15 models predict $J-K$ colors that change very little ($<0.1$ mag) with either mass or age in the mass range of 0.1-0.3 \msun\ and the age range 1-15 Myr.

The best fit light curve in the second scenario has a reduced-$\chi^2$ that is slightly higher than that of the first scenario (1.24 compared to 1.18, for the masked light curve). In this second case, the EB radii ratio is significantly smaller ($k = 0.6825\pm0.0081$), while the temperature ratio is $T_\mathrm{eff,2}/T_\mathrm{eff,1} \sim$ 1.22, from $J=2.187\pm0.015$. The component luminosities are such that $L_1=L_2=0.018\pm0.004$ \lsun, and $L_3=0.036\pm0.007$ \lsun. Since the ``secondary'' and tertiary have similar NIRC2 colors, we will assume they have equal temperatures in order to calculate the EB radii. The implied radii are then $R_1, R_2, R_3 = 0.746\pm0.082, 0.506\pm0.055, 0.712\pm0.078$ \rsun, respectively. The primary effective temperature in this case is $T_\mathrm{eff,1}=2440\pm60$ K. This scenario is also somewhat difficult to imagine, given that it implies the ``primary'' is $\sim$500 K cooler than the secondary, but with a radius that is $\sim$50\% larger due to the fact that the spectroscopic flux ratios provide a strong constraint that the EB components have nearly equal luminosities. We also explored fits with lower levels of contamination, and note that the $\chi^2_\mathrm{red}$ of these fits increased monotonically with lower third light values.

Ultimately, we adopt parameters assuming the first scenario ($L_\mathrm{EB}<L_3$), using the HH15 temperature for an M5 star for the ``secondary'', and allowing the third light parameter to be free. The best-fit orbital parameters and their uncertainties, derived from 1,000 MC simulations with \jktebop, as well as derived parameters for all three components are presented in Table~\ref{tab:epic608table}. We do not attempt to finely characterize the tertiary, due to the numerous intermediate assumptions required in doing so. We note that the most robust information we have for the third component is that it has a similar brightness to the unresolved EB in $J$ and $K$, has a similar $J-K$ color, and is undetected in all epochs of our optical spectra. 

We consider an alternative explanation for the simultaneous presence of a bright companion in the NIRC2 AO imaging and non-detection of a distinct third component in the HIRES spectra. If the closely projected source discovered in the AO images is not associated but instead a background M-giant, it may have a similar brightness and colors in the near-IR but be too faint to contribute significantly to the optical HIRES spectra. There are two primary difficulties in accepting this scenario: 1) at such a small projected separation ($\sim$0.1''), the probability that the source is unassociated is finite but low, and 2) the light curve modeling is highly suggestive that there is significant third light in the \Kepler\ bandpass, which is primarily optical but does extend to $\sim$0.9 \micron.

We stress that unquantifiable uncertainties remain for the EB parameters of \starC, and that the quoted uncertainties are merely formal errors. In particular, the masses are highly uncertain due to the fact that only one component has RVs that are well fit by the model. As illustrated above, uncertainties in the radii-related parameters on the order of a few to tens of percent may also remain due to faulty assumptions regarding the precise optical third light value. However, we again emphasize that there is compelling evidence that the sum of the RV semi-amplitudes is $<$60 \kms, which at the period implied by the light curve implies a total system mass $<$0.1 \msun, placing the components firmly in the brown dwarf mass regime. Furthermore, if the masses are indeed as low as $\sim$20 \mjup, and if the tertiary is in fact associated, this system constitutes a unique and intriguing comparison to the population of brown dwarfs and high mass giant planets on wide orbits (tens of AU) that are routinely imaged around young, mostly early-type stars.  

Comparing \starC\ with the compilation of pre-MS EBs and SBs presented in \cite{ismailov2014}, we note that independent of our difficulties above in determining the component parameters, this young system has the highest eccentricity for any pre-MS EB/SB system with a period below 10 days. However, the high eccentricity must also be considered in the context of the potential hierarchical triple nature of the system.

\subsection{EPIC 203476597}
\label{subsec:star597}

This system is comprised of a late-G type primary, with a likely mid-K-type secondary in a close circular orbit of period 1.4 d. The low inclination indicates grazing eclipses (and thus a poorly constrained radius ratio), and there is some evidence that the system is semi-detached.

A 5-pixel aperture produced the highest quality $K2$ photometry, which was selected for further correction. The raw light curve exhibits both $\sim$3.5\% primary and $\sim$2\% secondary eclipses with period 1.44 days (see Fig.~\ref{fig:lightcurve}). In addition, there is a roughly sinusoidal pattern due to rotation with a 3.21 day period. The 5-pixel aperture also contains a nearby star contributing $\sim$25\% of the total flux. Consequently, we subtracted the time-averaged flux from a 1.5-pixel aperture centered on the neighboring star. In principle, this subtraction removes dilution effects, restoring eclipses to their true depths. We note the eclipses became $\sim$1\% deeper after this subtraction. For each of the two stars, photometry was extracted from 1.5-pixel apertures to confirm EPIC 203476597 is the eclipsing source.

Twenty-two observations were discarded due to being flux outliers with quality flags indicating the spacecraft was in coarse pointing mode. The stellar variability was removed via four iterations of the cubic B-spline fit with 2-$\sigma$ outlier rejection upon each iteration. After removing the variability, an additional thirteen observations with flux levels 1-$\sigma$ \emph{above} the median were noted to be artifacts of the detrending procedure and were subsequently discarded. After the detrending procedure was applied, the phased light curve was divided into 100 bins. In each phase bin the mean and standard deviation were computed and 3-$\sigma$ outliers were identified, resulting in the exclusion of an additional 55 observations across the entire campaign.

According to \cite{rizzuto2015}, the primary is a G8 lithium-rich star with weak H$\alpha$ emission and a small amount of reddening ($A_V=1.3$ mag). Our spectrum is consistent with this type, though a K0 might be more appropriate for the line ratios seen in the HIRES spectrum.  In order to fit the SED a higher value of the reddening is found, $A_V\sim3.0$ mag, which can be lower if a small infrared excess is permitted beyond 2 $\mu$m, and drops to no less than 2 mag allowing for a spectral type as late as K2. The considerable extinction is consistent with the star's location towards optical nebulosity in the vicinity of $\rho$ Oph, just southwest of the main embedded cluster.  

The proper motion measurements reported in UCAC4 ($\mu_\alpha, \mu_\delta = -7.9, -19.9$ mas yr$^{-1}$) and PPMXL are consistent with membership in Upper Sco within $\chi^2 < 0.1-2.5$ (depending on which measurements and which mean cluster values are adopted).  Aiding membership confirmation is the detection of both \ion{Li}{1} absorption and weak H$\alpha$ emission.

As noted in Figure~\ref{fig:spectra}, there are obvious changes in H$\alpha$ line profiles among spectra of this eclipsing system. Examining the difference and ratio of the spectra reveals the change in the lines more clearly, and suggests that the secondary possesses weak H$\alpha$ and \ion{Ca}{2} triplet core emission as well as \ion{Li}{1} absorption.  It is challenging to infer an accurate spectral type from the spectral subtractions or ratios, but a mid-K (K2-K5) type is consistent with the data. Radial velocities obtained over seven epochs never exceeded 1.5 \kms\ in magnitude, despite extensive coverage in orbital phase. However, we do note that from epoch to epoch, a large velocity shift is noted in the H$\alpha$ emission component: a positive $10\pm2$ \kms\ shift was measured between the first and third epochs, separated by only a small phase difference, and a positive $98\pm2$ \kms\ shift between the first and second epochs which differed by almost 0.2 in phase. These measurements seem to suggest that while the primary shifted by only $\sim$2 \kms, from the first to second epochs, the secondary moved by $\approx$100 \kms. The RVs are presented in Table~\ref{table:rvs}, and the primary RV curve is presented in Fig.~\ref{fig:rv}. Though a fit is not found to the RVs, we can use the non-detection of orbital motion in the lines of the primary to place an upper limit on the mass ratio of $q\sim0.03$, which would place the secondary in the substellar mass regime. This scenario is seemingly inconsistent with inferences from the HIRES spectra.

One possible explanation for the non-detection of orbital motion greater than a few \kms\ in the primary is that the eclipses are due to an unassociated, young EB with low-mass components that are not detectable in the HIRES spectra, except for in H$\alpha$ emission, due to a low optical flux ratio with the G8-K0 star. In this scenario, dilution from the G8-KO star would dilute the eclipse depths of the EB and mimic a low inclination orbital configuration. More complete phase coverage is needed in the RV curve, but current observations imply a smaller mass ratio than the analysis below suggests.

At the model light curve fitting stage, we adopted a linear limb darkening law for both the primary and secondary. We fixed the limb darkening coefficients for both components to $u=0.7$, corresponding to the mean of all the tabulated values from \cite{sing2010} with $3.5 \leq \log{g} \leq 4.0$, 3500 K $\leq$ \teff\ $\leq$ 5500 K, and -0.1 $\leq$ [M/H] $\leq$ 0.1.

From the effective temperature of $5180\pm200$ K derived from the spectral type and HH15 calibration plus recommended error, we find from the photometry that $A_V=2.4\pm0.2$ mag and a $J$-band based system luminosity $\log{(L/L_\odot)}=0.13 \pm 0.11$ dex, where the error terms come from Monte Carlo sampling of the allowed error in temperature, but for luminosity are dominated by the uncertainty in the distance. The light curve modeling produces a luminosity ratio which is in good agreement with PARSEC model predictions of the luminosity ratio expected between 0.8 \msun\ and 1.4 \msun\ stars at 10 Myr.

\begin{figure}
\centering
\includegraphics[width=0.49\textwidth]{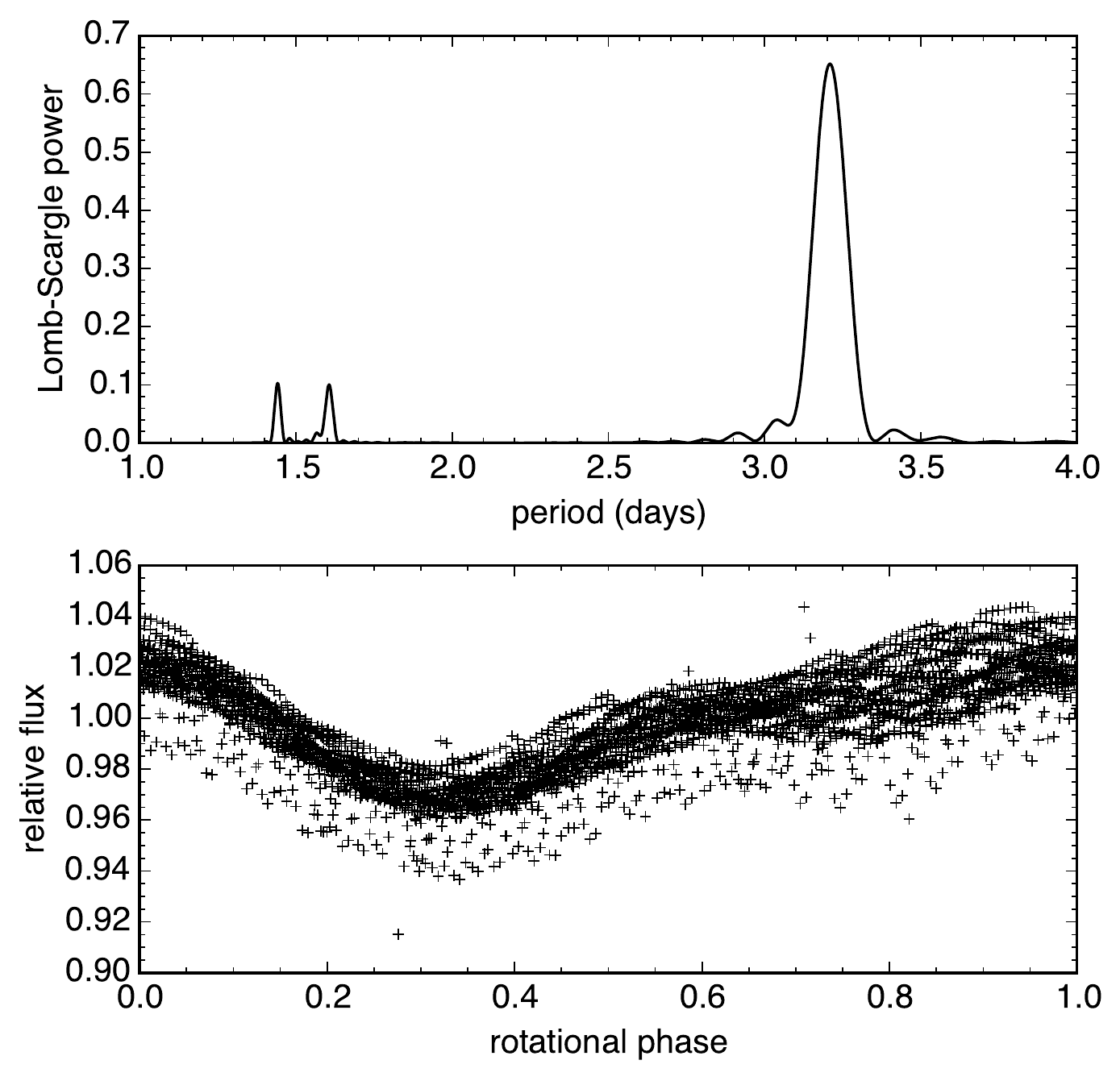}
\caption{\emph{Above}: A Lomb-Scargle periodogram analysis of the EPIC 203476597 raw light curve for 10,000 periods between 1 and 4 days, using the \texttt{lombscargle} routine in the \texttt{scipy.signal} {\sc python} package. The peak at 3.21 days is the rotational period of the primary, while the peaks at 1.4 and 1.6 days represent the orbital period and half the rotational period, respectively. \emph{Below}: The raw K2 light curve phase folded on the rotational period.}
\label{fig:epic597rotation}
\end{figure}

From the primary temperature and luminosity we calculated the primary radius to be $R_1 = 1.33 \pm 0.38$ \rsun. The primary radius can be better constrained, however, through combination of the rotational period and projected rotational velocity. The raw $K2$ light curve possesses variability due to rotational modulation of star spots. We performed a Lomb-Scargle periodogram analysis on the raw light curve for 10,000 periods between 1 and 4 days. The periodogram peak suggests a rotation period of $P_\mathrm{rot}$=3.21$\pm$0.12 days, where the uncertainty is estimated from the full-width half-maximum (FWHM) of a Gaussian fit to the oversampled periodogram peak. Figure~\ref{fig:epic597rotation} shows the periodogram described above along with the raw light curve phase folded on the rotational period. From the HIRES spectrum, we then measured a projected rotational velocity of $v\sin{i}$=25 $\pm$ 2 \kms. Combined with the rotational period, we calculate $R_1\sin{i}$ = 1.59$\pm$0.14 \rsun. Thus, we derive a lower limit of $R_1 > 1.45$ \rsun\ and for the range of inclinations favored by the light curve modeling, we find $R_1$=$1.72^{+0.17}_{-0.27}$ \rsun. 

Combining the primary radius with the best-fit ratio of radii we obtain a secondary radius of $R_2$ = 0.96 $\pm$ 0.27 \rsun, where the large uncertainty is due to the grazing nature of the eclipses. At presumed ages of 5-10 Myr (consistent with the star's location in both the mass-radius and temperature-luminosity planes), this range of radii corresponds to a late-K to mid-M type secondary according to the BHAC15 models, consistent with the inference from the HIRES spectrum. The expected RV semi-amplitude in the primary of this configuration is $\sim$ 70-100 \kms, or 130-160 \kms\ for the secondary depending on the broad ranges in plausible component masses. From the primary effective temperature and best-fit surface brightness ratio, $J$, we estimate the secondary temperature, $T_\mathrm{eff,2}=4490\pm60$ K. This value is consistent with a K3-K4 spectral type on the HH15 and PM13 scales.

The best-fit \jktebop\ model has an average fractional radius greater than 0.3, indicating the system may be semi-detached. \jktebop\ treats proximity effects (such as ellipsoidal modulation) in an approximate manner, and is best suited for modeling detached EBs. In such cases, the uncertainty in the derived radii may be as high as 5\% \citep{nz2004}, though the uncertainty in the ratio of radii we derived is much greater than 5\% due to the grazing configuration of the system. The best-fit model light curve and phase-folded $K2$ photometry are presented in Fig.~\ref{fig:bestfit}. Best-fit orbital parameters and their uncertainties, derived from 10,000 MC simulations with \jktebop, are presented in Table~\ref{tab:epic597table}.

From the primary radius and $R_1/a$ from the light curve solution, we computed the semi-major axis, $a = 5.8 \pm 0.9$ \rsun. However, this separation implies a total system mass which is lower than the presumed primary mass, given the period. We obtained lower and upper limits on the semi-major axis by considering the range in system mass corresponding to $M_\mathrm{tot} = M_1$ to $M_\mathrm{tot} = 2 M_1$, also accounting for the uncertainty on the mass. The corresponding range in semi-major axis is 5.8-7.9 \rsun, or $\sim 3-5.5$ times the primary radius. Note that this range assumes the model-dependent primary mass. \cite{kallrath2009} suggest that stars with radii greater than $\sim$10-15\% of their separation no longer meet the criterion for detachment, providing further support that this system is likely to be semi-detached.

The 2D parameter distributions resulting from the MC fit showed high degrees of correlation between the inclination, surface brightness ratio, and radii related parameters. Nevertheless, the total range in each of these parameters was deemed acceptable. We investigated an alternative solution, holding eccentricity fixed at zero, and fixing the period and ephemeris timebase. This solution yielded very similar results to those from allowing these same parameters to be free.

We note that the presence of both eclipses and a spot modulation pattern in the light curve of EPIC 203476597 may allow for determination of the direction of orbital motion. Eclipse timing variations induced by star spots, combined with measurements of the local slope in the variable light curve of the primary during eclipses, can allow one to distinguish between prograde and retrograde motion \citep{mazeh2015, holczer2015}.

\section{General Discussion}
\label{sec:general}

\begin{figure*}
\centering
\includegraphics[width=0.99\textwidth]{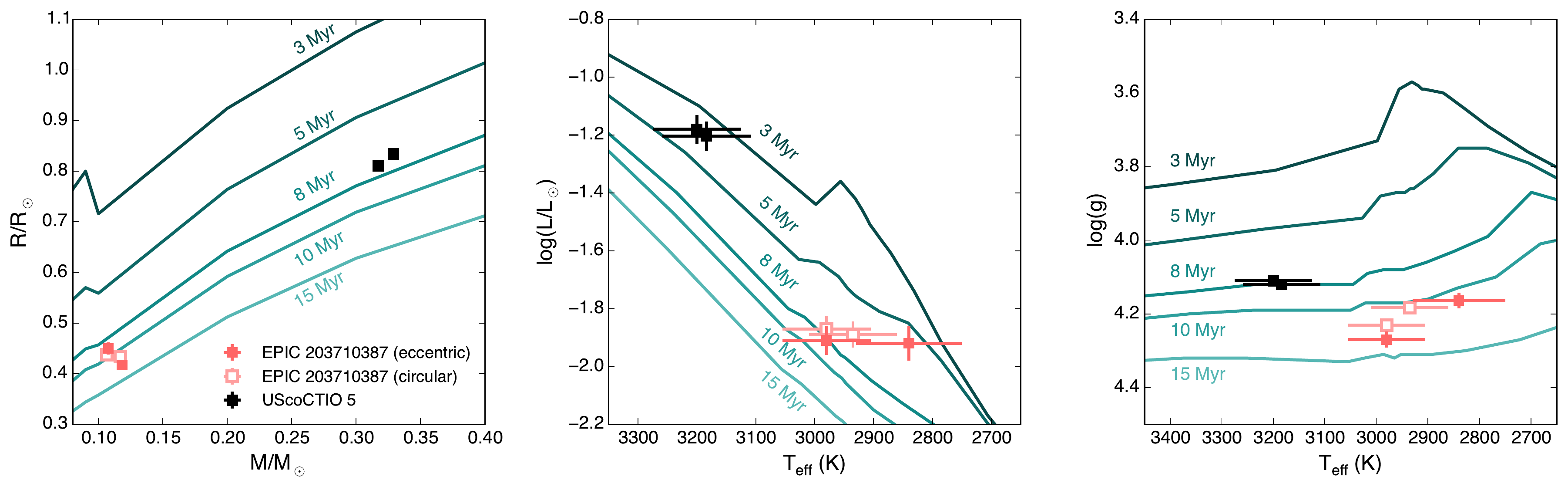}
\caption{BHAC15 isochrones showing an enhanced view of the mass-radius plane (left), as well as the \teff-$\log{(L/L_\odot)}$ (middle), and \teff-$\log{g}$ (right) planes. In each case the 3, 5, 8, 10, and 15 Myr isochrones are plotted, from darkest to lightest. The red points indicate the positions of both components of EPIC 203710387, while the black scatter points represent the components of UScoCTIO 5. The dark red shaded squares indicate the parameters of EPIC 203710387 from the eccentric orbit solution, while the light red open squares show the circular solution values. The components of UScoCTIO5 are assumed to have equal temperatures and luminosities, but are offset for clarity here. For both systems, the uncertainties in mass, radius, and $\log{g}$ are smaller than the points themselves.} 
\label{fig:threepanel}
\end{figure*}

The young eclipsing binaries identified here are a significant contribution to the pre-main sequence eclipsing binary population below 1 $M_\odot$. We have added two EB systems with M-type primary stars and one EB system with a late G primary star to the $<15-20$ EB systems already known at ages less than a few Myr (see \cite{ismailov2014, stassun2014} for compilations of pre-main sequence EBs and SBs).  Quantitative information for each system is provided in Tables~\ref{tab:epic387table}, ~\ref{tab:epic608table}, and ~\ref{tab:epic597table}.  

Our best determinations of the fundamental parameters for the components of the three systems are illustrated Figure~\ref{fig:isochrones}, in comparison to two other multiple systems in the same Upper Sco association with well-determined parameters.

Each of the three EBs considered in this work have periods $<5$ days -- even though the $K2$ data stream is sensitive to periods as long as 37 days (and up to 75 days if single eclipses are deemed significant). While the periods sample a range of parameter space occupied by other known low-mass EBs\footnote{http://www.astro.keele.ac.uk/$\sim$jkt/debcat/}, short period orbits are generally attributed to observational biases \citep{ribas2006}, which are not present in the case of the $K2$ data given its continuous cadence over the 75 days.  The period distribution of the newly discovered systems can be further compared to that for previously known pre-main sequence EBs with good orbital solutions, which span the range 2-14 days plus the recent 34 day system characterized by \cite{kraus2015}.   

Two of our systems (\starA\ and \starB) appear to be on highly circular orbits, while \starC\ has a non-negligible eccentricity of $e \approx 0.3$. This last system has the longest period at 4.5 d and is likely part of a hierarchical triple. However, according to \cite{zb1989}, circularization is expected to occur within $<$1 Myr for periods shorter than about 7 days. \citet{melo2001} seems to have a different view on the necessary timescales, and as noted by \cite{zahn2008}, the pre-MS circularization timescale is a topic of ongoing research. 

\starB\ constitutes the lowest {\it stellar} mass double-lined EB discovered to date.  With masses between the $\sim0.17 M_\odot + 0.18 M_\odot$ JW 380 pair of stars and the $\sim0.04 M_\odot + 0.06 M_\odot$ 2MASS J0535-0546 pair of brown dwarfs, both systems located in Orion, the older $\sim0.12 M_\odot + 0.11 M_\odot$ \starB\ pair in Upper Sco provides a critical anchor near the substellar boundary for pre-MS evolution models.   
\subsection{Comparing EPIC 203710387 and UScoCTIO 5}
\label{subsec:compare}
With very few low-mass, pre-MS EBs currently known, \starB\ and UScoCTIO 5, which as double-lined EB systems both have fundamentally determined masses and radii measured to $\lesssim 3\%$ precision, are extremely valuable for testing both model predictions and empirical relations. Interestingly, \starB\ is a slightly lower-mass analog to UScoCTIO 5 \citep[][see also the current Appendix]{kraus2015} in that it has a mass ratio close to 1 (though it has a much shorter orbital period, 2.8 d compared to 34 d). These systems are especially significant because at the lowest stellar masses, discrepancies between observations of eclipsing binaries and theoretical models are most prominent.

For example, main sequence EBs with M-type components have been observed to have radii that are 5-15\% larger than model predictions \citep{ribas2006}. Though, those authors do point out models seem to perform better below $\sim$0.30-0.35 \msun\ (near the limit between fully convective stars and those with radiative cores). Magnetic activity is one possible explanation invoked to account for the inflated radii of low-mass EBs. In principle, this is a testable prediction since the model-observation discrepancies should become larger at shorter periods due to the facts that (1) at short periods the rotational and orbital periods are expected to be synchronized and (2) activity is expected to increase with increasing rotational velocity \citep{feiden2015}.

For pre-MS evolution, starspots have also been advanced as a means of producing inflated radii for low-mass stars. Recently, \cite{somers2015} studied the effect of starspots on pre-MS evolution for stars of 0.1-1.2 \msun. They found that pre-MS models accounting for starspots leads to radii that are enhanced by up to 10\%, consistent with observations of active EBs. Spotted stars also have a decreased luminosity and \teff, leading to systematic underestimation of both masses (by a factor of 2) and ages (by factors of 2-10) derived from evolutionary models that do not take spots into account.

\cite{kraus2015} used USco-CTIO 5 to test various pre-main-sequence evolutionary models. For an assumed cluster age of 11 Myr, those authors found BHAC15 and several other models under-predict the fundamentally determined radius at the fundamentally determined mass (with Padova models working in the opposite direction).  For the more traditional cluster age of 5 Myr, the models over-predict the radius.  The results are consistent in the older age scenario with the so-called ``radius inflation" found among many main sequence eclipsing binary systems. 

We also find for the components of \starB\ that for the canonical 3-5 Myr age, the models significantly over-predict the fundamentally determined radii. However, we find that for an assumed cluster age of 10-11 Myr, the BHAC15 models quite accurately predict the radii at the masses of the components of \starB. Thus, if the older age is assumed accurate, we find no evidence for radius inflation in this lower mass, shorter period analog to UScoCTIO 5.

In comparing the two systems, we also noted significant temperature discrepancies for what is reportedly only a 0.5 subclass difference in spectral type. \cite{kraus2015} determined an M4.5$\pm$0.5 spectral type for UScoCTIO 5 based on comparison of  a low-resolution spectrum with field M dwarf spectra, simultaneously constraining spectral type and extinction. While this result is consistent with the M4 spectral type for UScoCTIO 5 originally reported by \cite{ardila2000}, there is evidence favoring an earlier type. \cite{reiners2005} found that discrepancies between the dynamically measured system mass and masses predicted by models could be rectified by considering a spectral type that is half a subclass or more earlier than M4.

Indeed, the ``geometric'' temperature derived by \cite{kraus2015} for both components of UScoCTIO 5 (from the sum of the radii, total system luminosity, and assuming equal-luminosity components) is $T_\mathrm{eff, geom} = 3235^{+160}_{-200}$ K, which is slightly higher than the empirical temperatures of young M4 stars on both the HH15 ($T_\mathrm{eff}=3190$ K) and PM13 ($T_\mathrm{eff}=3160$ K) scales. By comparison, the effective temperatures for the components of \starB\, based on the total system luminosity, luminosity ratio, and the radii (assumed to be equal), are more in line with an M6 type for both the primary and secondary on the HH15 scale. In other words, the components of \starB\ are $\sim$100 K cooler than the predicted M5 temperature on the HH15 scale, but in good agreement with the analogous PM13 prediction.

If we assume equal-luminosity components of \starB\ and calculate the ``geometric'' effective temperature implied by the total system luminosity and the sum of the radii, as \cite{kraus2015} did, we obtain $T_\mathrm{eff,geom} \approx$ 2410 $\pm$ 120 K, or more than 500 K cooler than the empirical temperature of a young M5 star. This discrepancy becomes larger if we assume the radii have been underestimated. In fact, assuming the M5 spectral type and temperature are correct, then the system luminosity implies a sum of radii of $R_1+R_2=0.568 \pm 0.057$ \rsun, or component radii of only $\sim$0.284 \rsun.

One possible explanation for this discrepancy is starspots. The ``geometric'' temperature assumes the measured luminosity is the intrinsic luminosity. However, for a spotted star the measured luminosity is actually the product of the intrinsic luminosity and the factor $(1-\beta)$, where $\beta$ is the equivalent spot covering fraction, which \cite{jackson2014} suggest may be as high as $\sim$0.35-0.51 for M-type pre-MS stars.

Thus, assuming the measured luminosity is the intrinsic luminosity leads to an anomalously low temperature when holding the radius fixed, or conversely an erroneously small radius when the temperature is held fixed. We estimate the average spot covering fraction for the components of \starB\ from the ratio of the ``emitting'' surface area to the measured surface area, which implies an average spot covering fraction of $\sim$60\%.

In Appendix~\ref{app:usco5}, we present our independent analysis of UScoCTIO 5 from our own detrended $K2$ light curve combined with the radial velocities and spectroscopic flux ratios published in \cite{kraus2015}. We find masses that are consistent with K15, but radii that are significantly larger. Our revised parameters help somewhat to resolve the discrepancies noted above for the age of the system as determined in different theoretical planes. We find the system age to be consistent with $\sim$6 Myr.

\subsection{On the Age of Upper Scorpius}
\label{subsec:ageofsco}
Disagreement about the age of Upper Sco stems from different studies of distinct stellar populations, such that the more massive stars appear older ($\sim$ 10 Myr) and the less massive stars appear younger ($\sim$ 3-5 Myr). Several possible ``simple'' explanations for this observed discrepancy exist: (1) the evolutionary models are inadequate, and the degree to which they diverge from observations is mass-dependent (this explanation includes the failure of models to properly include magnetic fields and spot-related effects), (2) the binary fraction at low masses is underestimated such that isochrone ages for these stars are anomalously young, having neglected the companion's luminosity, (3) there is a genuine dispersion of roughly a few Myr in the ages of Upper Sco members, indicating extended star formation.\footnote{Evidence for luminosity spreads, potentially due to an age dispersion of several Myr, is well-documented in the pre-MS Orion Nebula Cluster \citep{hillenbrand1997, dario2010}.}

Notably, the age we find for \starB\ in the fundamental mass-radius plane is $\sim$10-11 Myr (using the circular orbit solution), older than the canonical 3-5 Myr age for Upper Sco. While large uncertainties remain for \starA\ (grazing and possibly semi-detached) and \starC\ (a triple system), starB\ and UScoCTIO 5 are well-characterized and provide reliable anchors with which to investigate the cluster age at the lowest stellar masses.

In Figure~\ref{fig:threepanel}, we show the positions of these two double-lined EBs in different planes with BHAC15 isochrones overplotted. In each plane, no single isochrone can match the observed parameters of both EBs. However, an even larger discrepancy becomes apparent when comparing the ages of a single EB system derived in different planes. For example, while the components of \starB\ rest near the 10 Myr isochrone in mass-radius space, the same stars suggest an age of $\sim$7$\pm$3 Myr in the temperature-luminosity plane. A similar trend is true of UScoCTIO 5, the components of which lie closest to the 8 Myr mass-radius isochrone, but appear more consistent with the canonical 3-5 Myr age in temperature-luminosity space. It is noteworthy that the \teff-$\log{(L/L_\odot)}$ ages for both systems are in broad agreement with the widely accepted cluster age of 3-5 Myr which is also based on H-R diagram (HRD) analyses of low-mass members. The ages in \teff-$\log{g}$ space for each system, however, are in closer agreement with those ages from the mass-radius plane.

As reviewed by \cite{kraus2015}, considering different sets of pre-MS models does not alleviate the discrepancies noted above. These results indicate that (1) no set of models is able to predict the ensemble of fundamental parameters for pre-MS stars in this mass range, and/or (2) there is a systematic bias in one or more observationally determined parameters. For example, if current empirical SpT-\teff\ or color-\teff\ scales systematically underestimate \teff\ by a couple hundred K, the HRD-derived ages of the hundreds of low-mass members would shift closer to 10 Myr. However, in the \teff-$\log{g}$ plane it is not possible to shift the systems along the temperature axis in order to obtain a match to the canonical 3-5 Myr age. This indicates there is some minimal ``age'' spread if it is believed that radii contract and masses remain constant during pre-MS evolution.

The stellar bulk parameters of mass and radius for double-lined EBs are directly determined with exquisite precision based on firmly understood physics. Meanwhile, the photospheric parameters of \teff\ and luminosity are generally less well-determined. In principle, this suggests that ages derived for double-lined EBs in the mass-radius diagram should be considered more fundamental than HRD ages. However, few pre-MS EBs with well-determined radii exist, and so evolution models at these ages are uncalibrated.

If the masses and radii of EPIC 203710387 and UScoCTIO 5 are assumed accurate, then the models must overpredict the radii by $\sim$10-25\% for 5 Myr to be the true age. Interestingly, this implies stellar evolution models are not contracting quickly enough at these masses to match observations. Future iterations of stellar models at these masses will include magnetic fields and starspots, two phenomena which are intrinsically linked and both act to \emph{slow} contraction through inhibiting convection and decreasing the emergent flux \citep{feiden2015b}. Thus, as models evolve to include these effects, the discrepancy between the canonical age of 3-5 Myr and the ages implied for these two systems in the mass-radius plane will likely \emph{widen}. If current mass-radius isochrones are assumed correct, then these two double-lined EBs favor an older (8-10 Myr) age for Upper Sco.

\subsection{Coevality Within and Between Systems}
\label{subsec:coevality}

With multiple EBs in the same star forming region, and the additions of UScoCTIO 5 and HD 144548, it is possible to study the degree of coevality within individual systems (intra-coevality) and between distinct EBs (inter-coevality). \cite{stassun2014} found that, among PMS EBs in Orion, the components within a given EB appear significantly more coeval than do the EBs relative to one another. In other words, EBs in Orion display a higher degree of intra-coevality than inter-coevality. A possible explanation for this behavior could be genuine age dispersion in a presumably coeval population, as mentioned in \S~\ref{subsec:ageofsco}.

As mentioned in \S~\ref{subsec:triples}, the more massive component of the triple system HD 144548 appears to be several Myr younger than lower mass eclipsing pair. However, due to the difficulty of characterizing triple systems, it is possible that this apparent non-coevality within a single system is artificial in nature. Nevertheless, an empirical mass-radius isochrone at the age of Upper Sco is beginning to emerge from the components of EPIC 203710387, UScoCTIO 5, and the lower mass components of HD 144548 (see Fig.~\ref{fig:massradius}). Each of these systems have mass ratios close to 1, making it difficult to draw meaningful conclusions about the intra-coevality of any particular system. However, there is an interesting trend in which the higher mass EBs of Upper Sco appear \emph{younger} than their lower mass counterparts. 

From our comparison of EPIC 203710387 and UScoCTIO 5 (see Fig.~\ref{fig:threepanel}), it is also apparent that there is a higher degree of coevality within each of these systems than between them. Specifically, there is a $\lesssim$ 1 Myr discrepancy between the ages of the primary and secondary of EPIC 203710387 (in the circular orbit case) and similarly for the two components of UScoCTIO 5. However, despite the fact that both systems belong to the presumably coeval population in Upper Sco, EPIC 203710387 appears to be 3-4 Myr older than UScoCTIO 5. Preliminary results from modeling the pre-main-sequence evolution of low-mass stars indicate that including magnetic fields may significantly help to resolve the discrepancies noted above (G. Feiden, private communication). We note that the eccentric orbit solution for EPIC 203710387 leads to primary and secondary parameters that appear significantly less coeval relative to the circular solution in all of the parameter planes featured in Fig.~\ref{fig:threepanel}. However, we caution that this discrepancy is possibly due to a degeneracy between the temperature ratio and ratio of radii. More precise spectroscopic flux ratios may help to resolve this issue in the future.

\subsection{Chromospheric Activity Effects}
\label{subsec:activity}

The correlation between chromospheric activity and the temperatures, radii, and subsequently derived masses of main-sequence stars is well-established \citep[e.g.][]{lopezmorales2007}. Due to the paucity of pre-MS benchmark systems, however, the effect of activity on the fundamental parameters of pre-MS stars has not been rigorously tested.  For field age low-mass stars and brown dwarfs, \cite{stassun2012} derived an empirical relation to correct for such activity effects based on the H$\alpha$ equivalent width (see \S 2.1 of that work), among other measures.  Using our EW(H$\alpha$) measurements from the HIRES spectra (see Table~\ref{table:ews}), we calculated that the fractional change in the radii and temperatures for the two components of EPIC 203710387 are approximately:
\begin{align*}
\Delta R_1/R_1 &= 2.0 \pm 0.9\% \\
\Delta R_2/R_2 &= 2.7 \pm 1.3\% \\
\Delta T_\mathrm{eff,1}/T_\mathrm{eff,1} &= -1.3 \pm 0.4 \% \\
\Delta T_\mathrm{eff,2}/T_\mathrm{eff,2} &= -1.6 \pm 0.6 \%. 
\end{align*}
Thus, in agreement with the findings of \cite{stassun2014}, we determine that the effect of chromospheric activity on the temperatures, radii, and masses of EPIC 203710387 is small enough that it can not resolve the apparent discrepancies between the positions in the mass-radius diagram compared with the positions in $T_\mathrm{eff}$-$L$ space, as discussed in \S~\ref{subsec:compare} and illustrated in Fig.~\ref{fig:threepanel}. We note, however, that a temperature suppression of $\sim$2\% does help to partially resolve the discrepancies noted above, leading to inferred ages closer to those implied by the well-determined masses and radii.

We find for the eclipsing components of EPIC 203868608 that the corrections in the radii and temperatures due to activity are approximately $\Delta R/R \sim 4 \pm 2 \%$, and $\Delta T_\mathrm{eff}/T_\mathrm{eff} \sim 2 \pm 1 \%$. This level of temperature suppression could potentially explain the apparent reversal in \teff\ between the primary and secondary components, given that the primary is apparently more active. Similar behavior was observed in the other known eclipsing brown dwarf system discovered in Orion \citep{stassun2006, stassun2007}. We note, however, even larger uncertainties may remain in the parameters of this system due to its triple nature and the subsequent complexity of its analysis.

\subsection{Triple Systems in Upper Sco}
\label{subsec:triples}
\begin{figure*}
\centering
\includegraphics[width=0.95\textwidth]{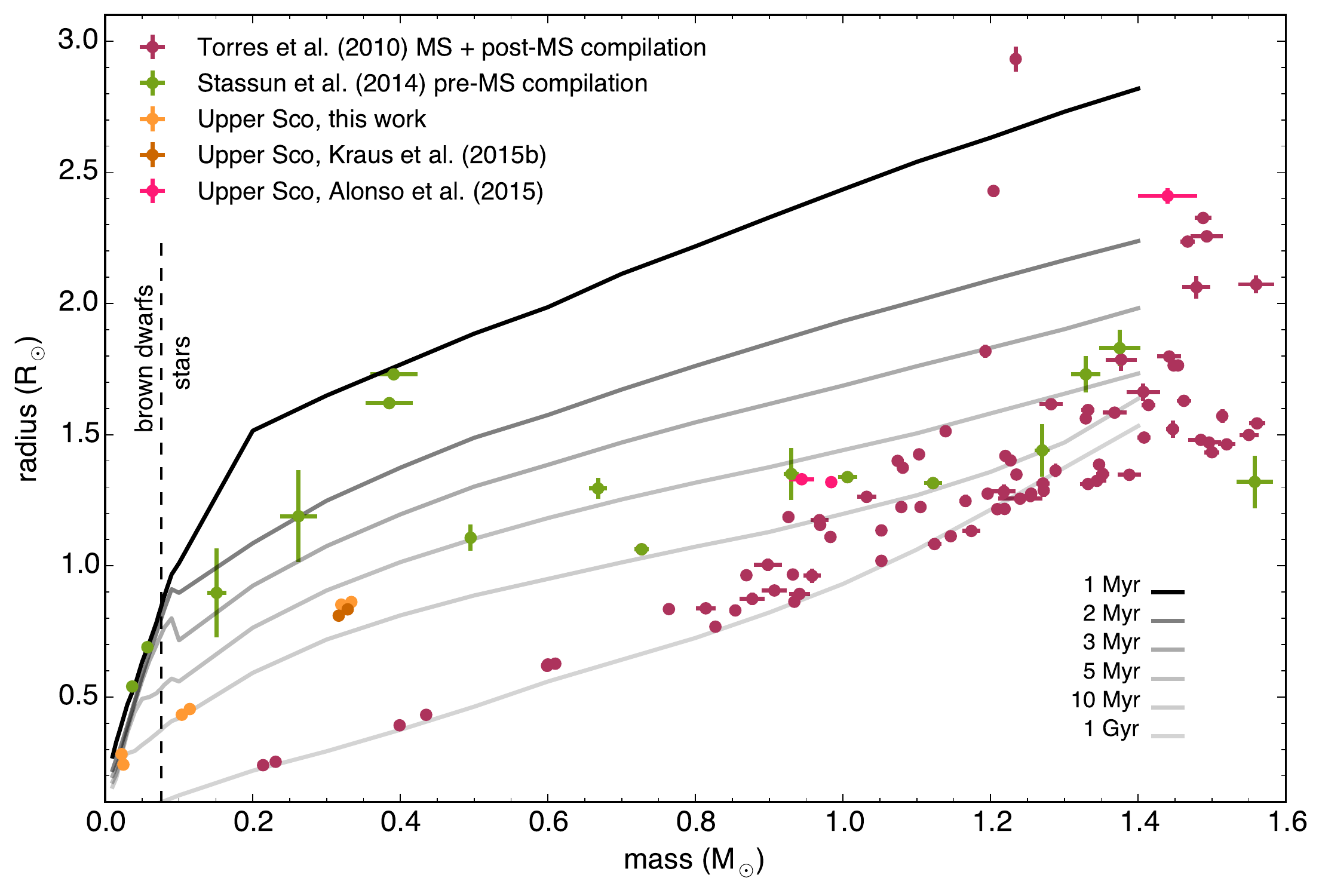}
\caption{BHAC15 isochrones in the mass-radius plane. Overplotted are compilations of double-lined EBs with fundamentally determined masses and radii, either in the pre-MS phase of evolution \citep{stassun2014} or MS/post-MS phases of evolution \citep{torres2010}. At a fixed mass, the radius evolves vertically downward in this diagram. We include recently characterized,  double-lined eclipsing members of Upper Sco. For UScoCTIO 5, first characterized by \cite{kraus2015}, we overplot our revised parameters. We additionally input small offsets to our derived parameters for EPIC 203710387 and EPIC 203868608 for visual clarity. We do not include the tertiary for this latter system since a fundamental determination of the mass and radius for that component was not possible. We stress that unquantifiable uncertainties remain for EPIC 203868608, but we include the EB components here for illustrative purposes. The pink points correspond to the triply eclipsing system HD 144548 \citep{alonso2015}.}
\label{fig:massradius}
\end{figure*}

In comparison with binaries, \cite{stassun2014} found that benchmark pre-MS triple systems have apparently corrupted properties in both the mass-radius and \teff-luminosity planes. Recently, \cite{alonso2015} characterized the young triply eclipsing system HD 144548 (EPIC 204506777) in Upper Sco. As seen in Figure~\ref{fig:massradius}, the system is composed of an eclipsing pair of $\sim$1.0 \msun\ stars, which in turn eclipse a $\sim$1.5 \msun\ tertiary host. While the less massive pair have masses and radii that are in broad agreement with the emerging empirical mass-radius isochrone from EPIC 203710387 and UScoCTIO 5 (i.e. located between the 5 and 10 Myr BHAC15 isochrones), the massive tertiary has a highly discrepant mass and radius suggestive of a 1-2 Myr age.

ScoPMS 20 is another triple system in Upper Sco, characterized by \cite{mace2012}. As seen in Figure~\ref{fig:isochrones}, this system also presents a challenge to the conventional notion of a coeval stellar population within Upper Sco. Only one component of ScoPMS 20 has a published mass and radius, which places it between 2-5 Myr in widely used pre-MS models. This is in contrast to EPIC 203710387, UScoCTIO 5, and the two less massive components of HD 144548, which are all suggestive of an age between 5-10 Myr according to BHAC15 mass-radius isochrones. The three components of ScoPMS 20 present a slightly more coherent picture in the temperature-luminosity plane, though the positions are suggestive of a somewhat younger age relative to the lower mass systems mentioned above.

Finally, for EPIC 203868608, the uncertainties intrinsic to the analysis of this system make it difficult to comment on the quality of derived parameters relative to binary counterparts. With our current knowledge, it is not possible to determine the radius of the tertiary, since there is no eclipse information and even if the orbit is highly inclined, the period implied by the separation is unfavorably long. Moreover, it is not yet clear whether the tertiary is indeed associated. Resolved spectroscopy of the closely projected companion could be used to test the scenario that it is a background M-giant. If the companion is associated, it is possible, however, that radial velocity time series over a sufficiently long time baseline could allow for dynamical mass measurements of all three components. The radii are uncertain due to the uncertain nature of the contamination in the $K2$ bandpass. Optical AO imaging of the system would provide a direct measurement of this quantity, and further test the background M-giant scenario described above. Furthermore, the EB components appear to be in a mass regime where the pre-MS evolutionary tracks are closely clustered and largely vertical (see Figure~\ref{fig:massradius}), implying that a broad range of radii are feasible for current mass determinations and the relatively uncertain age of Upper Sco.

\section{Conclusion}

We report the discovery of three new pre-MS EBs. Two systems (\starA\ and \starB) are secure members of the Upper Sco association, while the third (\starC) is certainly young but has a discrepant proper motion. All three systems are located in the southern part of the association, relatively close to but west of the $\rho$ Oph molecular cloud. 

The system \starB\ was observed to be double-lined, allowing  model-independent masses and radii to be measured through combination of the light curve and the radial velocities.  With near-equal mass 0.12+0.11 $M_\odot$ components, it is the lowest-mass stellar double-lined EB discovered to date.\footnote{The pre-MS system reported by \cite{stassun2006, stassun2007} has lower mass components with masses just below the stellar-brown dwarf boundary.}  The mass measurements of both components have $\sim$2\% precision, while the radii were fixed as equal in order to obtain a reasonable solution for the ensemble of parameters. The positions of both components in the fundamental mass-radius plane are consistent with a $\sim$10 Myr age according to both the BHAC15 and \cite{siess2000} isochrones. 

We provide tentative evidence that the system \starC\ is an eclipsing system of $\sim$25+25 \mjup\ brown dwarfs in a potential hierarchical triple configuration with a wide M-type companion. If confirmed, this would be only the second double-lined eclipsing brown dwarf system discovered to date \citep[see][]{stassun2006, stassun2007}.  This system also constitutes the most eccentric pre-MS binary system having an orbital period $<$10 days \citep[c.f.][]{ismailov2014}, though a stellar-mass companion interior to 20 AU also contributes to the dynamical evolution of this system. Such a system presents a unique data point for studies concerning pre-MS circularization  timescales. The triple nature of the system also makes it interesting for investigations of dynamical effects in hierarchical triples such as the Kozai-Lidov mechanism.  This system is also significantly more luminous than \starB\ while sharing the same combined light spectral type. Follow-up studies, notably optical AO imaging and resolved near-IR spectroscopy could shed light on the nature of the AO companion. Additional radial velocities will also help to more accurately constrain the EB component masses and separation, and hence radii through combination of RVs and the light curve.

\starA\ has a roughly 1.4 \msun\ primary with a likely early-M to mid-K type secondary. The extremely short period suggests this system may be semi-detached, and there is possible evidence for ellipsoidal modulation in the raw $K2$ light curve. If ellipsoidal modulation is recovered from the light curve, re-analysis with software suitable for semi-detached EBs could produce a highly precise mass ratio for this system \citep{wilson1994}. Follow-up infrared spectroscopy, where the flux ratio is more favorable relative to optical, could reveal secondary lines and allow for dynamical mass measurements and directly measured radii. The positions of the primary in both the mass-radius and temperature-luminosity planes are consistent with an age of $\sim$10 Myr, though we note that the mass determination is model-dependent and the large parameter uncertainties do admit ages $<$5 Myr.

We have characterized the components of the three EB systems presented here based on the information available, acknowledging that future spectroscopic studies will greatly refine the parameters. These three newly identified EB systems, in addition to the recently fully characterized \citep{kraus2015} UScoCTIO 5 system, are valuable assets for constraining pre-MS evolutionary models at the age of Upper Sco. 
   
\acknowledgments
We thank the referee for many helpful comments, which greatly improved the quality and rigor of this work. We thank Ian Crossfield for lending his Python MCMC wrapper for the \jktebop\ orbit fitting code and for consultation regarding its use, which aided our analysis.  We thank Erik Petigura for helpful discussions regarding eclipse model fitting practices. We thank Howard Isaacson, Geoff Marcy, Erik Petigura and the CPS group for acquiring additional HIRES spectra and providing MAGIQ snapshots for the three systems. We thank Christoph Baranec for acquiring Keck/NIRC2 images of \starA\ and \starC, and Brendan Bowler for performing the reductions of these data. We thank Jessie Christiansen for early advice regarding \Kepler\ telescope data products, photometry techniques, and detrending strategies. We thank Avi Shporer for suggesting the possibility of measuring the direction of orbital motion for EPIC 203476597. We thank Jonathan Swift for lending his transit analysis routines, which aided our analysis. We thank Ross Fredella for his assistance in creating Figure 1.

The material presented herein is based upon work supported in 2015 by the National Science Foundation Graduate Research Fellowship under Grant No. DGE1144469. T. J. D. gratefully acknowledges research activities support from France C\'{o}rdova through the 
Neugebauer Scholarship. This research was partially supported by an appointment to the NASA Postdoctoral Program at the Ames Research Center, administered by Oak Ridge Associated Universities through a contract with NASA. This research has made use of the NASA Exoplanet Archive, which is operated by the California Institute of Technology, under contract with the National Aeronautics and Space Administration under the Exoplanet Exploration Program. This research has also made use of the SIMBAD database and VizieR catalog access tool, operated at CDS, Strasbourg, France, and NASA's ADS and IPAC/IRSA services. Some of the data presented in this paper were obtained from the Mikulski Archive for Space Telescopes (MAST). STScI is operated by the Association of Universities for Research in Astronomy, Inc., under NASA contract NAS5-26555. Support for MAST for non-HST data is provided by the NASA Office of Space Science via grant NNX09AF08G and by other grants and contracts. This paper includes data collected by the Kepler mission. Funding for the Kepler mission is provided by the NASA Science Mission directorate. Some of the data presented herein were obtained at the W.M. Keck Observatory, which is operated as a scientific partnership among the California Institute of Technology, the University of California and the National Aeronautics and Space Administration. The Observatory was made possible by the generous financial support of the W.M. Keck Foundation. The authors wish to recognize and acknowledge the very significant cultural role and reverence that the summit of Mauna Kea has always had within the indigenous Hawaiian community.  We are most fortunate to have the opportunity to conduct observations from this mountain.

\clearpage

\appendix

\newcommand{\sbratio}{0.825}
\newcommand{\esbratio}{0.065}
\newcommand{\sumradii}{0.1700}
\newcommand{\esumradii}{0.0021}
\newcommand{\ratradii}{1.077}
\newcommand{\eratradii}{0.045}
\newcommand{\inclination}{82.84}
\newcommand{\einclination}{0.10}
\newcommand{\ecosw}{-0.00298}
\newcommand{\eecosw}{0.00029}
\newcommand{\esinw}{-0.0153}
\newcommand{\eesinw}{0.0091}
\newcommand{\period}{2.808849}
\newcommand{\eperiod}{0.000024}
\newcommand{\tzero}{894.71388}
\newcommand{\etzero}{0.00051}
\newcommand{\rada}{0.0818}
\newcommand{\erada}{0.0020}
\newcommand{\radb}{0.0882}
\newcommand{\eradb}{0.0021}
\newcommand{\lightratio}{0.957}
\newcommand{\elightratio}{0.034}
\newcommand{\ecc}{0.0156}
\newcommand{\eecc}{0.0087}
\newcommand{\omegap}{259.0}
\newcommand{\eomegap}{9.3}
\newcommand{\bpri}{1.547}
\newcommand{\ebpri}{0.047}
\newcommand{\bsec}{1.500}
\newcommand{\ebsec}{0.021}
\newcommand{\rchisq}{1.060}
\newcommand{\erchisq}{0.043}
\newcommand{\rmslc}{6.76}
\newcommand{\rvprim}{43.43}
\newcommand{\ervprim}{0.60}
\newcommand{\rvsec}{47.77}
\newcommand{\ervsec}{0.43}
\newcommand{\rvsys}{-3.38}
\newcommand{\ervsys}{0.22}
\newcommand{\rmsprv}{0.75}
\newcommand{\rmssrv}{0.68}
\newcommand{\sma}{5.100} 
\newcommand{\esma}{0.043} 
\newcommand{\mrat}{0.909} 
\newcommand{\emrat}{0.014} 
\newcommand{\mpri}{0.1183} 
\newcommand{\empri}{0.0028} 
\newcommand{\msec}{0.1076} 
\newcommand{\emsec}{0.0031} 
\newcommand{\rpri}{0.417} 
\newcommand{\erpri}{0.010} 
\newcommand{\rsec}{0.450} 
\newcommand{\ersec}{0.012} 
\newcommand{\logga}{4.270} 
\newcommand{\elogga}{0.022} 
\newcommand{\loggb}{4.164} 
\newcommand{\eloggb}{0.021} 
\newcommand{\rhoa}{1.63} 
\newcommand{\erhoa}{0.12} 
\newcommand{\rhob}{1.184} 
\newcommand{\erhob}{0.088} 

\LongTables
\begin{deluxetable*}{lrlll}[bt]
\tabletypesize{\scriptsize}
\tablecaption{  System Parameters of EPIC 203710387 \label{tab:epic387table}}
\tablewidth{0.9\textwidth}
\tablehead{
\colhead{Parameter} & \colhead{Symbol or} & \colhead{Value} & \colhead{Units} & \colhead{Source} \\
& \colhead{Prefix} & & & 
} 

\startdata
\multicolumn{5}{c}{\emph{Identifying Information}} \\
\hline \\
Right ascension & $\alpha$ J2000. & 16:16:30.681 & hh:mm:ss & \cite{roeser2010} \\
Declination & $\delta$ J2000. & -25:12:20.20 & dd:mm:ss & \cite{roeser2010} \\
  $K2$ ID & EPIC & 203710387 & & Huber \& Bryson 2015\\
  2 Micron All Sky Survey ID & 2MASS &  J16163068-2512201 & & \cite{cutri2003}  \\
  Wide-field Infrared Survey Explorer ID & AllWISE & J161630.66-251220.3 & & \cite{cutri2014} \\ 
 UKIRT Infrared Deep Sky Survey ID & UGCS & J161630.67-251220.2 & & \cite{lawrence2013} \\
\\
\hline \\
\multicolumn{5}{c}{\emph{Photometric Properties}} \\
\hline \\
 & $J$   & 12.932 $\pm$ 0.023 & mag & 2MASS\\
 & $H$   & 12.277 $\pm$ 0.024 & mag & 2MASS\\
 & $K_s$ & 11.907 $\pm$ 0.023 & mag & 2MASS\\
 & $W1$  & 11.748 $\pm$ 0.023 & mag & AllWISE \\
 & $W2$  & 11.483 $\pm$ 0.022 & mag & AllWISE \\
 & $W3$  & $>$11.559          & mag & AllWISE; S/N $<2$\\
 & $W4$  & $>$8.846           & mag & AllWISE; S/N $<2$ \\
 & KEPMAG & 14.268             & mag & $K2$ EPIC \\
\hline\\
\multicolumn{5}{c}{\emph{Best-Fitting Adjusted Parameters (Eccentric Solution)}} \\
\hline \\
Orbital period & $P$ & \period\ $\pm$ \eperiod & days & this work \\
Ephemeris timebase - 2456000 & $T_0$ & \tzero\ $\pm$ \etzero & BJD & this work \\
Surface brightness ratio & $J$ & \sbratio\ $\pm$ \esbratio & & this work \\
Sum of fractional radii & $(R_1+R_2)/a$ & \sumradii\ $\pm$ \esumradii & & this work \\
Ratio of radii & $k$ & \ratradii\ $\pm$ \eratradii & & this work \\
Orbital inclination & $i$ & \inclination\ $\pm$ \einclination & deg & this work \\
Combined eccentricity, periastron longitude & $e\cos\omega$ & \ecosw\ $\pm$ \eecosw & & this work \\
Combined eccentricity, periastron longitude & $e\sin\omega$ & \esinw\ $\pm$ \eesinw & & this work \\
Primary radial velocity amplitude & $K_1$ & \rvprim\ $\pm$ \ervprim & km s$^{-1}$ & this work \\
Secondary radial velocity amplitude & $K_2$ & \rvsec\ $\pm$ \ervsec & km s$^{-1}$ & this work \\
Systemic radial velocity & $\gamma$ & \rvsys\ $\pm$ \ervsys & km s$^{-1}$ & this work \\
Mass ratio & $q$ & \mrat\ $\pm$ \emrat & & this work \\
Reduced chi-squared of light curve fit & $\chi^2_\text{red}$ & \rchisq &  & this work \\
RMS of best fit light curve residuals & & \rmslc & ppt & this work \\
Reduced chi-squared of primary RV fit & $\chi^2_\text{red}$ & 1.363  &  & this work \\
RMS of primary RV residuals & & \rmsprv & km s$^{-1}$ & this work \\
Reduced chi-squared of secondary RV fit & $\chi^2_\text{red}$ & 0.522 &  & this work \\
RMS of secondary RV residuals & & \rmssrv & km s$^{-1}$ & this work \\
\hline\\
\multicolumn{5}{c}{\emph{Best-Fitting Derived Parameters (Eccentric Solution)}}\\
\hline \\
Orbital semi-major axis & $a$ & \sma\ $\pm$ \esma & \rsun & this work \\
Fractional radius of primary & $R_\text{1}/a$ & \rada $\pm$ \erada & & this work \\
Fractional radius of secondary & $R_\text{2}/a$ & \radb $\pm$ \eradb & & this work \\
Luminosity ratio & $L_2/L_1$ & \lightratio\ $\pm$ \elightratio & & this work \\
Primary mass & $M_1$ & \mpri\ $\pm$ \empri & \msun & this work \\
Secondary mass & $M_2$ & \msec\ $\pm$ \emsec & \msun & this work \\
Primary radius & $R_1$ & \rpri\ $\pm$ \erpri & \rsun & this work \\
Secondary radius & $R_2$ & \rsec\ $\pm$ \ersec & \rsun & this work \\
Primary surface gravity & $\log g_1$ & \logga\ $\pm$ \elogga & cgs & this work \\
Secondary surface gravity & $\log g_2$ & \loggb\ $\pm$ \eloggb & cgs & this work \\
Primary mean density & $\rho_1$ & \rhoa\ $\pm$ \erhoa & $\rho_\odot$ & this work \\
Secondary mean density & $\rho_2$ & \rhob\ $\pm$ \erhob & $\rho_\odot$ & this work \\
Impact parameter of primary eclipse & $b_1$ & \bpri\ $\pm$ \ebpri & & this work \\
Impact parameter of secondary eclipse & $b_2$ & \bsec\ $\pm$ \ebsec & & this work \\
Eccentricity & $e$ & \ecc\ $\pm$ \eecc & & this work \\
Periastron longitude & $\omega$ & \omegap\ $\pm$ \eomegap & deg & this work \\
\hline\\
\multicolumn{5}{c}{\emph{Best-Fitting Adjusted Parameters (Circular Solution)}}\\
\hline \\
Orbital period & $P$ & 2.808862 $\pm$ 0.000024 & days & this work \\
Ephemeris timebase - 2456000 & $T_0$ & 894.71117 $\pm$  0.00043 & BJD & this work \\
Surface brightness ratio & $J$ & 0.940 $\pm$ 0.014 & & this work \\
Sum of fractional radii & $(R_1+R_2)/a$ & 0.1715 $\pm$ 0.0021 & & this work \\
Ratio of radii & $k$ & 1.009 $\pm$ 0.017 & & this work \\
Orbital inclination & $i$ & 82.76 $\pm$ 0.10 & deg & this work \\
Primary radial velocity amplitude & $K_1$ & 43.28 $\pm$ 0.52 & km s$^{-1}$ & this work \\
Secondary radial velocity amplitude & $K_2$ & 47.55 $\pm$ 0.57 & km s$^{-1}$ & this work \\
Systemic radial velocity & $\gamma$ & -3.26 $\pm$ 0.23 & km s$^{-1}$ & this work \\
Mass ratio & $q$ & 0.910 $\pm$ 0.015 & & this work \\
Reduced chi-squared of light curve fit & $\chi^2_\text{red}$ & 1.173 &  & this work \\
RMS of best fit light curve residuals & & 7.08 & ppt & this work \\
Reduced chi-squared of primary RV fit & $\chi^2_\text{red}$ & 1.423 &  & this work \\
RMS of primary RV residuals & & 0.80 & km s$^{-1}$ & this work \\
Reduced chi-squared of secondary RV fit & $\chi^2_\text{red}$ & 0.766 &  & this work \\
RMS of secondary RV residuals & & 0.85 & km s$^{-1}$ & this work \\

\hline\\
\multicolumn{5}{c}{\emph{Best-Fitting Derived Parameters (Circular Solution)}}\\
\hline \\
Orbital semi-major axis & $a$ & 5.044 $\pm$ 0.037 & \rsun & this work \\
Fractional radius of primary & $R_\text{1}/a$ & 0.0854 $\pm$ 0.0013 & & this work \\
Fractional radius of secondary & $R_\text{2}/a$ & 0.0861 $\pm$ 0.0013 & & this work \\
Luminosity ratio & $L_2/L_1$ & \lightratio\ $\pm$ \elightratio & & this work \\
Primary mass & $M_1$ & 0.1169 $\pm$ 0.0031 & \msun & this work \\
Secondary mass & $M_2$ & 0.1065 $\pm$ 0.0027 & \msun & this work \\
Primary radius & $R_1$ & 0.4338 $\pm$ 0.0071 & \rsun & this work \\
Secondary radius & $R_2$ & 0.4377 $\pm$ 0.0080 & \rsun & this work \\
Primary surface gravity & $\log g_1$ & 4.231 $\pm$ 0.013 & cgs & this work \\
Secondary surface gravity & $\log g_2$ & 4.183 $\pm$ 0.013 & cgs & this work \\
Primary mean density & $\rho_1$ & 1.433 $\pm$ 0.062 & $\rho_\odot$ & this work \\
Secondary mean density & $\rho_2$ & 1.270 $\pm$ 0.058 & $\rho_\odot$ & this work\\
Impact parameter of primary eclipse & $b_1$ & 1.4694 $\pm$ 0.0030 & & this work \\
Impact parameter of secondary eclipse & $b_2$ & 1.4694 $\pm$ 0.0030 & & this work \\
\hline \\
\multicolumn{5}{c}{\emph{Final Adopted Stellar Parameters}} \\
\hline \\
Primary spectral type & SpT$_1$ & M4.5-M5 & & this work, spectroscopy \\
Secondary spectral type & SpT$_2$ & M4.5-M5 & & this work, spectroscopy \\
Extinction & $A_V$ & 1.22 $\pm$ 0.31 & mag & this work, SpT, photometry\\
Bolometric luminosity & $\log{(L_\mathrm{bol}/L_\odot)}$ & -1.64 $\pm$ 0.08 & dex & this work, SpT, photometry, $A_V$, $d$\\
Primary luminosity & $L_1$ & 0.0124 $\pm$ 0.0014 & \lsun & this work, $T_\mathrm{eff,1}$, $R_1$ \\
Secondary luminosity & $L_2$ & 0.0119 $\pm$ 0.0016 & \lsun & this work, $T_\mathrm{eff,2}$, $R_2$ \\
Orbital semi-major axis & $a$ & \sma\ $\pm$ \esma & \rsun & this work, fundamental determination \\
Primary mass & $M_1$ & \mpri\ $\pm$ \empri & \msun & this work, fundamental determination \\
Secondary mass & $M_2$ & \msec\ $\pm$ \emsec & \msun & this work, fundamental determination \\
Primary radius & $R_1$ & \rpri\ $\pm$ \erpri & \rsun & this work, fundamental determination \\
Secondary radius & $R_2$ & \rsec\ $\pm$ \ersec & \rsun & this work, fundamental determination \\
Primary surface gravity & $\log g_1$ & \logga\ $\pm$ \elogga & cgs & this work, $M_1$, $R_1$ \\
Secondary surface gravity & $\log g_2$ & \loggb\ $\pm$ \eloggb & cgs & this work, $M_2$, $R_2$ \\
Primary mean density & $\rho_1$ & \rhoa\ $\pm$ \erhoa & $\rho_\odot$ & this work, $M_1$, $R_1$ \\
Secondary mean density & $\rho_2$ & \rhob\ $\pm$ \erhob & $\rho_\odot$ & this work, $M_2$, $R_2$ \\
Primary effective temperature & $T_\mathrm{eff,1}$ & 2980 $\pm$ 75 & K & this work, SpT, HH15 \\
Secondary effective temperature & $T_\mathrm{eff,2}$ & 2840 $\pm$ 90 & K & this work, $J$, $T_\mathrm{eff,1}$ \\
Primary age & $\tau_1$ & $11.6\pm0.4$ & Myr & this work \\
Secondary age & $\tau_2$ & $9.9\pm0.3$ & Myr & this work 

\enddata
\tablecomments{Best-fit orbital parameters and their uncertainties resulting from 10,000 Monte Carlo simulations with \jktebop\ in the circular case and 5,000 simulations in the eccentric case. The $\chi^2_\mathrm{red}$ values quoted above were computed over the light curve with out-of-eclipse observations removed. Both the primary and secondary ages are determined from interpolation of the MC distributions in mass and radius between the BHAC15 isochrones.}

\end{deluxetable*}


\newcommand{\ra}{16:17:18.992}
\newcommand{\dec}{-24:37:18.75}
\newcommand{\epicid}{203868608}
\newcommand{\twomassid}{J16171898-2437186}
\newcommand{\wiseid}{J161718.97-243718.9}
\newcommand{\ukidssid}{J161718.97-243718.7}

\newcommand{\pmra}{1.3}
\newcommand{\pmdec}{-19.6}
\newcommand{\pmref}{\cite{roeser2010}}
\newcommand{\av}{2.04 $\pm$ 0.31}
\newcommand{\avref}{this work\footnote{assumes SpT and photometry}}

\newcommand{\twomassj}{11.858 $\pm$ 0.026}
\newcommand{\twomassh}{11.137 $\pm$ 0.024}
\newcommand{\twomassk}{10.760 $\pm$ 0.021}
\newcommand{\wiseone}{10.535 $\pm$ 0.023}
\newcommand{\wisetwo}{10.286 $\pm$ 0.020}
\newcommand{\wisethree}{10.150 $\pm$ 0.078}
\newcommand{\wisefour}{8.638 $\pm$ 0.416}
\newcommand{\ukidssh}{11.166 $\pm$ 0.001}
\newcommand{\ukidssk}{10.720 $\pm$ 0.001}
\newcommand{\kepmag}{13.324}

\renewcommand{\sbratio}{1.223}
\renewcommand{\esbratio}{0.066}
\renewcommand{\sumradii}{0.12930}
\renewcommand{\esumradii}{0.00073}
\renewcommand{\ratradii}{0.904}
\renewcommand{\eratradii}{0.026}
\renewcommand{\inclination}{87.77}
\renewcommand{\einclination}{0.18}
\renewcommand{\ecosw}{-0.05377}
\renewcommand{\eecosw}{0.00011}
\renewcommand{\esinw}{0.3182}
\renewcommand{\eesinw}{0.0042}
\newcommand{\thirdlight}{0.684}
\newcommand{\ethirdlight}{0.016}
\renewcommand{\period}{4.541710}
\renewcommand{\eperiod}{0.000019}
\renewcommand{\tzero}{896.19699}
\renewcommand{\etzero}{0.00019}
\renewcommand{\rada}{0.0679}
\renewcommand{\erada}{0.0012}
\renewcommand{\radb}{0.06138}
\renewcommand{\eradb}{0.00071}
\renewcommand{\lightratio}{0.999}
\renewcommand{\elightratio}{0.027}
\renewcommand{\ecc}{0.3227}
\renewcommand{\eecc}{0.0042}
\renewcommand{\omegap}{99.59}
\renewcommand{\eomegap}{0.14}
\renewcommand{\bpri}{0.389}
\renewcommand{\ebpri}{0.024}
\renewcommand{\bsec}{0.751}
\renewcommand{\ebsec}{0.051}
\renewcommand{\rchisq}{1.185}
\renewcommand{\rmslc}{2.48}
\renewcommand{\rvprim}{25.74}
\renewcommand{\ervprim}{0.31}
\renewcommand{\rvsec}{23.17}
\renewcommand{\ervsec}{0.28}
\renewcommand{\rvsys}{-7.62}
\renewcommand{\ervsys}{0.25}
\renewcommand{\rmsprv}{6.21}
\renewcommand{\rmssrv}{0.63}
\renewcommand{\sma}{4.157} 
\renewcommand{\esma}{0.025} 
\renewcommand{\mrat}{1.111} 
\renewcommand{\emrat}{0.024} 
\renewcommand{\mpri}{0.02216} 
\renewcommand{\empri}{0.00045} 
\renewcommand{\msec}{0.02462} 
\renewcommand{\emsec}{0.00055} 
\renewcommand{\rpri}{0.2823} 
\renewcommand{\erpri}{0.0051} 
\renewcommand{\rsec}{0.2551} 
\renewcommand{\ersec}{0.0036} 
\renewcommand{\logga}{3.882} 
\renewcommand{\elogga}{0.017} 
\renewcommand{\loggb}{4.015} 
\renewcommand{\eloggb}{0.011} 
\renewcommand{\rhoa}{0.985} 
\renewcommand{\erhoa}{0.054} 
\renewcommand{\rhob}{1.482} 
\renewcommand{\erhob}{0.052} 

\clearpage
\LongTables
\begin{deluxetable*}{lrlll}[bt]
\tabletypesize{\scriptsize}
\tablecaption{  System Parameters of EPIC 203868608 \label{tab:epic608table}}
\tablewidth{0.9\textwidth}
\tablehead{
\colhead{Parameter} & \colhead{Symbol or} & \colhead{Value} & \colhead{Units} & \colhead{Source} \\
& \colhead{Prefix} & & & 
}

\startdata
\multicolumn{5}{c}{\emph{Identifying Information}} \\
\hline \\
Right ascension & $\alpha$ J2000. & \ra & hh:mm:ss & \cite{roeser2010} \\
Declination & $\delta$ J2000. & \dec & dd:mm:ss & \cite{roeser2010} \\
  $K2$ ID & EPIC & \epicid & & Huber \& Bryson 2015\\
  2 Micron All Sky Survey ID & 2MASS &  \twomassid & & \cite{cutri2003}  \\
  Wide-field Infrared Survey Explorer ID & AllWISE & \wiseid & & \cite{cutri2014} \\ 
 UKIRT Infrared Deep Sky Survey ID & UGCS & \ukidssid & & \cite{lawrence2013} \\
\\
\hline \\
\multicolumn{5}{c}{\emph{Photometric Properties}} \\
\hline \\
 & $J$    & \twomassj  & mag & 2MASS\\
 & $H$    & \twomassh  & mag & 2MASS\\
 & $K_s$  & \twomassk  & mag & 2MASS\\
 & $W1$   & \wiseone   & mag & AllWISE \\
 & $W2$   & \wisetwo   & mag & AllWISE \\
 & $W3$   & \wisethree & mag & AllWISE \\
 & $W4$   & \wisefour  & mag & AllWISE \\
 & KEPMAG  & \kepmag    & mag & $K2$ EPIC \\
\\
\hline\\
\multicolumn{5}{c}{\emph{Best-Fitting Adjusted Parameters}} \\
\hline \\
Orbital period & $P$ & \period $\pm$ \eperiod\ & days & this work \\
Ephemeris timebase - 2456000 & $T_0$ & \tzero\ $\pm$ \etzero & BJD & this work \\
Surface brightness ratio & $J$ & \sbratio\ $\pm$ \esbratio & & this work \\
Sum of fractional radii & $(R_1+R_2)/a$ & \sumradii\ $\pm$ \esumradii & & this work \\
Ratio of radii & $k$ & \ratradii\ $\pm$ \eratradii\ & & this work \\
Third light & $l_3$ & \thirdlight\ $\pm$ \ethirdlight\ & $L_\mathrm{tot}$ & this work \\
Orbital inclination & $i$ & \inclination\ $\pm$ \einclination & deg & this work \\
Combined eccentricity, periastron longitude & $e\cos\omega$ & \ecosw\ $\pm$ \eecosw & & this work \\
Combined eccentricity, periastron longitude & $e\sin\omega$ & \esinw\ $\pm$ \eesinw & & this work \\
Primary radial velocity amplitude & $K_1$ & \rvprim\ $\pm$ \ervprim & km s$^{-1}$ & this work \\
Secondary radial velocity amplitude & $K_2$ & \rvsec\ $\pm$ \ervsec & km s$^{-1}$ & this work \\
Systemic radial velocity & $\gamma$ & \rvsys\ $\pm$ \ervsys & km s$^{-1}$ & this work \\
Mass ratio & $q$ & \mrat\ $\pm$ \emrat & & this work \\ 
\hline\\
\multicolumn{5}{c}{\emph{Best-Fitting Derived Parameters}}\\
\hline \\
Orbital semi-major axis & $a$ & \sma\ $\pm$ \esma & \rsun & this work \\
Fractional radius of primary & $R_\text{1}/a$ & \rada\ $\pm$ \erada & & this work \\
Fractional radius of secondary & $R_\text{2}/a$ & \radb\ $\pm$ \eradb & & this work \\
Luminosity ratio & $L_2/L_1$ & \lightratio\ $\pm$ \elightratio & & this work \\
Primary mass & $M_1$ & \mpri\ $\pm$ \empri & \msun & this work \\
Secondary mass & $M_2$ & \msec\ $\pm$ \emsec & \msun & this work \\
Primary radius & $R_1$ & \rpri\ $\pm$ \erpri & \rsun & this work \\
Secondary radius & $R_2$ & \rsec\ $\pm$ \ersec & \rsun & this work \\
Primary surface gravity & $\log g_1$ & \logga\ $\pm$ \elogga & cgs & this work\\
Secondary surface gravity & $\log g_2$ & \loggb\ $\pm$ \eloggb & cgs & this work\\
Primary mean density & $\rho_1$ & \rhoa\ $\pm$ \erhoa & $\rho_\odot$ & this work \\
Secondary mean density & $\rho_2$ & \rhob\ $\pm$ \erhob & $\rho_\odot$ & this work \\
Impact parameter of primary eclipse & $b_1$ & \bpri\ $\pm$ \ebpri & & this work \\
Impact parameter of secondary eclipse & $b_2$ & \bsec\ $\pm$ \ebsec & & this work \\
Eccentricity & $e$ & \ecc\ $\pm$ \eecc & & this work \\
Periastron longitude & $\omega$ & \omegap\ $\pm$ \eomegap & deg & this work \\
Reduced chi-squared of light curve fit & $\chi^2_\text{red}$ & \rchisq &  & this work \\
RMS of best fit light curve residuals & & \rmslc & ppt & this work \\
Reduced chi-squared of primary RV fit & $\chi^2_\text{red}$ & 47.04 &  & this work \\
RMS of primary RV residuals & & 6.21 & km s$^{-1}$ & this work \\
Reduced chi-squared of secondary RV fit & $\chi^2_\text{red}$ & 1.381 &  & this work \\
RMS of secondary RV residuals & & 0.63 & km s$^{-1}$ & this work \\

\hline \\
\multicolumn{5}{c}{\emph{Other Adopted Stellar Parameters}} \\
\hline \\
Spectral Type & SpT & M5$\pm$0.5 & & this work, spectroscopy \\
Extinction & $A_V$ & \av & mag & this work, SpT, photometry \\
Bolometric luminosity & $\log{(L_\mathrm{bol}/L_\odot)}$ & -1.14 $\pm$ 0.08 & dex & this work, SpT, photometry, $A_V$, $d$\\
Primary effective temperature & $T_\mathrm{eff,1}$ & 2830 $\pm$ 80 & K & this work, $T_\mathrm{eff,2}$, $J$\\
Secondary effective temperature & $T_\mathrm{eff,2}$ & 2980 $\pm$ 75 & K & this work, SpT, HH15

\enddata
\tablecomments{Best-fit orbital parameters and their uncertainties are the result of 1,000 Monte Carlo simulations with \jktebop. The $\chi^2_\mathrm{red}$ quoted above was computed for the light curve with out-of-eclipse observations removed, and is reduced to 1.052 over the full light curve.}

\end{deluxetable*}


\renewcommand{\ra}{16:25:57.915}
\renewcommand{\dec}{-26:00:37.35}
\renewcommand{\epicid}{203476597}
\renewcommand{\twomassid}{J16255790-2600374}
\renewcommand{\wiseid}{J162557.90-260037.5}
\renewcommand{\ukidssid}{J162557.91-260037.5}

\renewcommand{\pmra}{-7.9}
\renewcommand{\pmdec}{-19.9}
\renewcommand{\pmref}{\cite{zacharias2013}}
\newcommand{\ewha}{-0.14 $\pm$ 0.08}
\newcommand{\ewharef}{\cite{rizzuto2015}}
\newcommand{\ewli}{0.31 $\pm$ 0.06}
\newcommand{\ewliref}{\cite{rizzuto2015}}
\renewcommand{\av}{1.3}
\renewcommand{\avref}{\cite{rizzuto2015}}

\renewcommand{\twomassj}{9.575 $\pm$ 0.024}
\renewcommand{\twomassh}{8.841 $\pm$ 0.044}
\renewcommand{\twomassk}{8.535 $\pm$ 0.021}
\renewcommand{\wiseone}{8.161 $\pm$ 0.019}
\renewcommand{\wisetwo}{8.130 $\pm$ 0.017}
\renewcommand{\wisethree}{8.230 $\pm$ 0.023}
\renewcommand{\wisefour}{7.757 $\pm$ 0.172}
\newcommand{\ukidssz}{11.112 $\pm$ 0.001}
\newcommand{\ukidssy}{11.005 $\pm$ 0.001}
\newcommand{\ukidssj}{10.485 $\pm$ 0.001}
\renewcommand{\ukidssh}{11.971 $\pm$ 0.001}
\renewcommand{\ukidssk}{10.219 $\pm$ 0.000}
\renewcommand{\kepmag}{9.575}

\renewcommand{\sbratio}{0.563}
\renewcommand{\esbratio}{0.032}
\renewcommand{\sumradii}{0.462}
\renewcommand{\esumradii}{0.014}
\renewcommand{\ratradii}{0.56}
\renewcommand{\eratradii}{0.13}
\renewcommand{\inclination}{67.5}
\renewcommand{\einclination}{1.2}
\renewcommand{\ecosw}{-0.00035}
\renewcommand{\eecosw}{0.00017}
\renewcommand{\esinw}{0.0029}
\renewcommand{\eesinw}{0.0030}
\renewcommand{\period}{1.4408031}
\renewcommand{\eperiod}{0.0000050}
\renewcommand{\tzero}{894.37787}
\renewcommand{\etzero}{0.00016}
\renewcommand{\rvprim}{0.10}
\renewcommand{\ervprim}{0.27}
\renewcommand{\rvsys}{-0.67}
\renewcommand{\ervsys}{0.22}
\renewcommand{\rada}{0.296}
\renewcommand{\erada}{0.016}
\renewcommand{\radb}{0.166}
\renewcommand{\eradb}{0.030}
\renewcommand{\lightratio}{0.176}
\renewcommand{\elightratio}{0.090}
\renewcommand{\ecc}{0.0029}
\renewcommand{\eecc}{0.0027}
\renewcommand{\omegap}{97.0}
\renewcommand{\eomegap}{14.1}
\renewcommand{\bpri}{1.29}
\renewcommand{\ebpri}{0.13}
\renewcommand{\bsec}{1.30}
\renewcommand{\ebsec}{0.13}
\renewcommand{\rchisq}{1.46}
\renewcommand{\erchisq}{}
\renewcommand{\rmslc}{1.06}

\clearpage
\LongTables
\begin{deluxetable*}{lrrll}[bt]
\tabletypesize{\scriptsize}
\tablecaption{  System Parameters of EPIC 203476597 \label{tab:epic597table}}
\tablewidth{0.9\textwidth}
\tablehead{
\colhead{Parameter} & \colhead{Symbol} & \colhead{Value} & \colhead{Units} & \colhead{Source} \\
& \colhead{or Prefix} & & & 
}

\startdata
\multicolumn{5}{c}{\emph{Identifying Information}} \\
\hline \\
Right ascension & $\alpha$ J2000. & \ra & hh:mm:ss & \cite{zacharias2013} \\
Declination & $\delta$ J2000. & \dec & dd:mm:ss & \cite{zacharias2013} \\
  $K2$ ID & EPIC & \epicid & & Huber \& Bryson (2015) \\
  2 Micron All Sky Survey ID & 2MASS &  \twomassid &  & \cite{cutri2003}  \\
  Wide-field Infrared Survey Explorer ID & AllWISE & \wiseid & &  \cite{cutri2014} \\ 
 UKIRT Infrared Deep Sky Survey ID & UGCS & \ukidssid &  & \cite{lawrence2013} \\
\\
\hline \\
\multicolumn{5}{c}{\emph{Photometric Properties}} \\
\hline \\
 & $J$    & \twomassj  & mag & 2MASS\\
 & $H$    & \twomassh  & mag & 2MASS\\
 & $K_s$  & \twomassk  & mag & 2MASS\\
 & $W1$   & \wiseone   & mag & AllWISE \\
 & $W2$   & \wisetwo   & mag & AllWISE \\
 & $W3$   & \wisethree & mag & AllWISE \\
 & $W4$   & \wisefour  & mag & AllWISE \\
 & KEPMAG  & \kepmag    & mag & $K2$ EPIC \\
\\
\hline\\
\multicolumn{5}{c}{\emph{Best-Fitting Adjusted Parameters}} \\
\hline \\
Orbital period & $P$ & \period\ $\pm$ \eperiod & days & this work \\
Ephemeris timebase - 2456000 & $T_0$ & \tzero\ $\pm$ \etzero & BJD & this work \\
Surface brightness ratio & $J$ & \sbratio\ $\pm$ \esbratio & & this work \\
Sum of fractional radii & $(R_1+R_2)/a$ & \sumradii\ $\pm$ \esumradii & & this work \\
Ratio of radii & $k$ & \ratradii\ $\pm$ \eratradii & & this work \\
Orbital inclination & $i$ & \inclination\ $\pm$ \einclination & deg & this work \\
Combined eccentricity, periastron longitude & $e\cos\omega$ & \ecosw\ $\pm$ \eecosw & & this work \\
Combined eccentricity, periastron longitude & $e\sin\omega$ & \esinw\ $\pm$ \eesinw & & this work \\
Primary radial velocity amplitude & $K_1$ & \rvprim\ $\pm$ \ervprim & km s$^{-1}$ & this work \\
Systemic radial velocity & $\gamma$ & \rvsys\ $\pm$ \ervsys & km s$^{-1}$ & this work \\
Reduced chi-squared of light curve fit & $\chi^2_\text{red}$ & \rchisq &  & this work\\
RMS of best fit light curve residuals & & \rmslc & ppt & this work \\
\hline\\
\multicolumn{5}{c}{\emph{Best-Fitting Derived Parameters}} \\
\hline \\
Fractional radius of primary & $R_\text{1}/a$ & \rada\ $\pm$ \erada & & this work \\
Fractional radius of secondary & $R_\text{2}/a$ & \radb\ $\pm$ \eradb & & this work \\
Luminosity ratio & $L_2/L_1$ & \lightratio\ $\pm$ \elightratio & & this work \\
Impact parameter of primary eclipse & $b_1$ & \bpri\ $\pm$ \ebpri & & this work \\
Impact parameter of secondary eclipse & $b_2$ & \bsec\ $\pm$ \ebsec & & this work \\
Eccentricity & $e$ & \ecc\ $\pm$ \eecc & & this work \\
Periastron longitude & $\omega$ & \omegap\ $\pm$ \eomegap & deg & this work \\
\hline \\
\multicolumn{5}{c}{\emph{Other Adopted Stellar Parameters}} \\
\hline \\
Spectral Type & SpT & G8-K0  & & this work \\
Extinction & $A_V$ & 2.42 $\pm$ 0.52 & mag & this work \\
Primary effective temperature & $T_\mathrm{eff,1}$ & 5180 $\pm$ 200 & K & this work, SpT, HH15 \\
Secondary effective temperature & $T_\mathrm{eff,2}$ & 4490 $\pm$ 60 & K & this work, $T_\mathrm{eff,1}$, $J$ \\
Bolometric luminosity & $\log{(L_\mathrm{bol}/L_\odot)}$ & 0.13 $\pm$ 0.11 & dex & this work, SpT, photometry, $A_V$, $d$\\
Primary luminosity & $L_1$ & 1.15$\pm$0.66 & \lsun & this work, $L_\mathrm{bol}$, $L_2/L_1$\\
Secondary luminosity & $L_2$ & 0.20$\pm$0.12 & \lsun & this work, $L_\mathrm{bol}$, $L_2/L_1$\\
Primary rotation period & $P_\mathrm{rot}$ & 3.21$\pm0.12$ & days & this work\\
Projected rotational velocity & $v\sin{i}$ & 25$\pm$2 & \kms & this work\\
Primary radius & $R_1$ & 1.72$^{+0.17}_{-0.27}$ & \rsun & this work, $P_\mathrm{rot}$, $v\sin{i}$ \\
Secondary radius & $R_2$ & 0.96 $\pm$ 0.27 & \rsun & this work, $R_1$, $k$\\
Primary mass & $M_1$ & 1.41$\pm$0.17 & \msun & this work, $T_\mathrm{eff,1}$, $R_1$, PARSEC\\
Secondary mass & $M_2$ & 0.84$\pm$0.12 & \msun & this work, $T_\mathrm{eff,2}$, $R_2$, PARSEC\\
Primary age & $\tau_1$ & 6.6$^{+2.4}_{-3.6}$ & Myr & this work, $T_\mathrm{eff,1}$, $R_1$, PARSEC\\
\enddata
\tablecomments{Best-fit orbital parameters and their uncertainties are the result of 1,000 Monte Carlo simulations with \jktebop. We note that the $\chi^2_\mathrm{red}$ of the model light curve fit was computed with out-of-eclipse observations removed, and becomes $<$1 when computed over the entire light curve. }
\end{deluxetable*}

\newpage

\section{Revised Parameters for UScoCTIO 5} \label{app:usco5}

\begin{figure}
\centering
\includegraphics[width=0.95\textwidth]{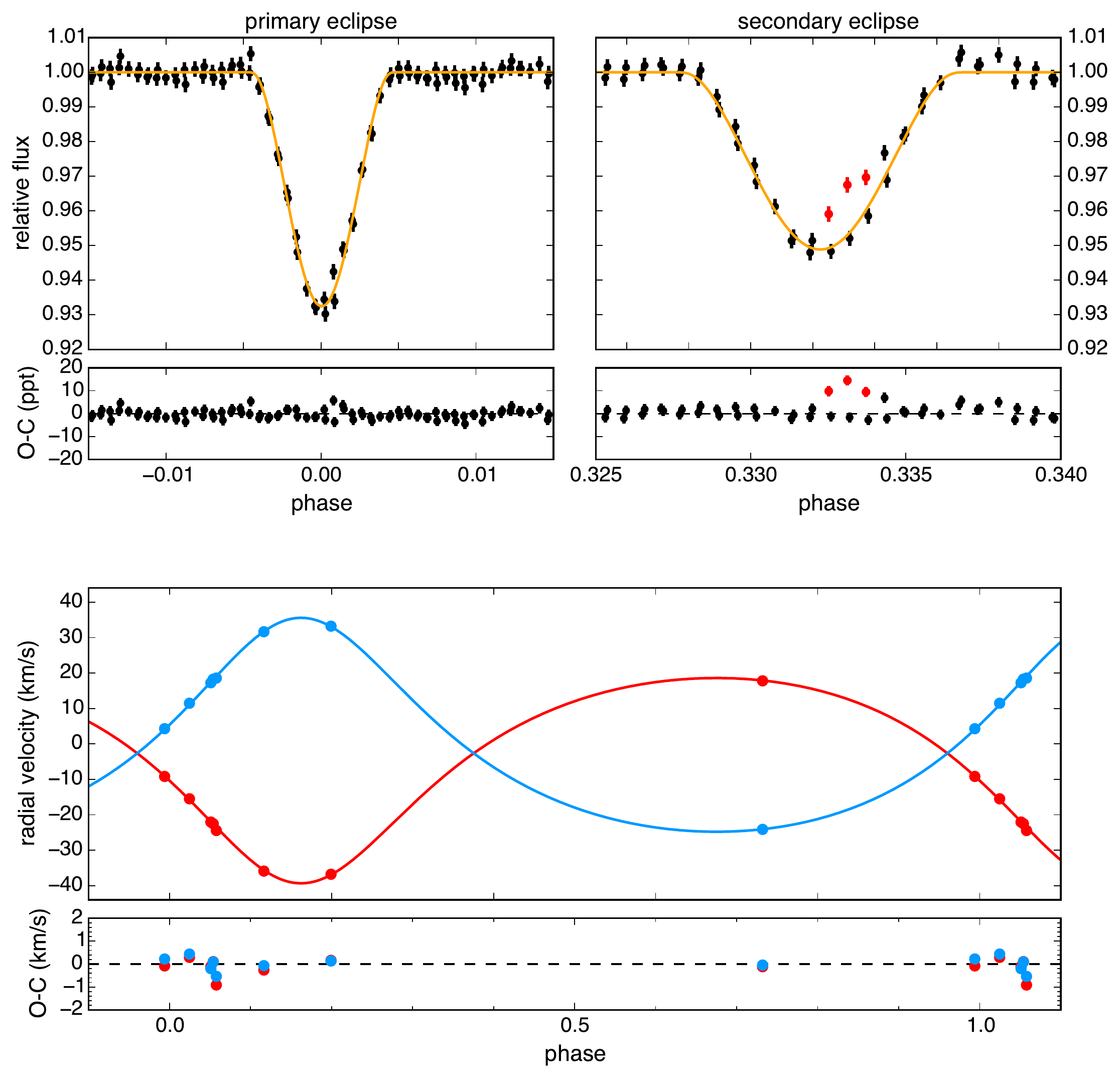}
\caption{Best-fit \jktebop\ models to our detrended $K2$ light curve for UScoCTIO 5 (top panels) and the radial velocities published by K15 (bottom panel). The three red points in secondary eclipse (upper right), were excluded from the fitting procedure. These points, as well as others, were also excluded in the K15 analysis.}
\label{fig:usco5fit}
\end{figure}

We independently characterized the double-lined eclipsing system UScoCTIO 5 (EPIC 205030103), using our own detrended $K2$ light curve and the radial velocities and spectroscopic flux ratios published in Table 1 of \cite{kraus2015}, hereafter K15. We used a 3-pixel aperture to extract photometry from the target pixel files, and detrended the raw light curve using our procedure described in \S~\ref{subsec:detrending}. Following the same approach for EPIC 203710387 (which has similar spectral type components), we assumed a linear limb-darkening law for both components, fixing the coefficient $u$ to 0.888 for each. Consistent with our analysis above, we also account for the \Kepler\ long cadence integration time through numerical integration of the models at ten points across intervals of 1766 seconds.  Because we used a larger photometric aperture than employed in K15, we investigated the possibility of contamination by allowing third light as a free parameter in a \jktebop\ trial fit. The resulting best-fit third light value was consistent with zero and so this parameter was fixed at zero for the final fit. We also excluded three clear outliers during secondary eclipse in our final fit, which we note were also excluded from the fitting procedure in K15. These outliers are possibly systematic artifacts intrinsic to the data, or perhaps related to the detrending procedure, though given the fact that they are seen in independently detrended light curves it is also possible there is modulation of the eclipse morphology due to star spots. As initial parameter estimates for the final fit, we used the best-fit values found by K15 as input for 10,000 MC simulations with \jktebop. We present our newly derived best-fit parameters in Table~\ref{tab:usco5table}, and show the best-fit models to the photometry and radial velocities in Figure~\ref{fig:usco5fit}.

We find component masses that are consistent at the $2-\sigma$ level with those published in K15. However, we find radii that are discrepant with those in K15 at the $>4.5-\sigma$ level for the primary and the $7-\sigma$ level for the secondary (where we take the uncertainties from K15 as 1-$\sigma$ in a given parameter), such that our radii are larger. The implication of this finding is that the UScoCTIO 5 component positions in the mass-radius plane are consistent with an age slightly younger than the age implied by the K15 parameters.

Combining the bolometric luminosity from K15, with our radii determinations and the luminosity ratio implied by our best-fit model, we find temperatures of $T_\mathrm{eff,1}$=3180$\pm$180 K and $T_\mathrm{eff,2}$=3140$\pm$180 K, consistent with K15. The large temperature uncertainties are dominated by the large uncertainty in the bolometric luminosity. These temperatures are consistent with matching M4 spectral types on both the HH15 and PM13 empirical scales. For completeness, we also investigated the possible effect of chromospheric activity on the temperatures and radii for UScoCTIO 5. Using the empirical relations of \cite{stassun2012} and the H$\alpha$ equivalent widths published in K15, we estimate that activity may account for an additional $\sim$1\% change in the temperatures and radii for this system.

Most notably, our revised radii for the components of UScoCTIO 5 help to resolve the discrepancies noted in \S~\ref{subsec:compare}, in that the K15 parameters produce an age in the H-R diagram that is nearly a factor of two younger than the age implied in the mass-radius plane, when using BHAC15 models. Using our newly derived parameters, there is better agreement in the age of the system as derived in the mass-radius, \teff-log($L/L_\odot$), and \teff-log($g$) planes with BHAC15 models.

\renewcommand{\period}{34.000703}
\renewcommand{\eperiod}{0.000089}
\renewcommand{\tzero}{909.25110}
\renewcommand{\etzero}{0.00085}
\renewcommand{\sbratio}{0.955}
\renewcommand{\esbratio}{0.035}
\renewcommand{\sumradii}{0.04473}
\renewcommand{\esumradii}{0.00048}
\renewcommand{\ratradii}{0.989}
\renewcommand{\eratradii}{0.018}
\renewcommand{\inclination}{87.880}
\renewcommand{\einclination}{0.025}
\renewcommand{\ecosw}{-0.266564}
\renewcommand{\eecosw}{0.000071}
\renewcommand{\esinw}{0.0191}
\renewcommand{\eesinw}{0.0031}
\renewcommand{\rvprim}{28.962}
\renewcommand{\ervprim}{0.090}
\renewcommand{\rvsec}{30.185}
\renewcommand{\ervsec}{0.085}
\renewcommand{\rvsys}{-2.651}
\renewcommand{\ervsys}{0.043}
\renewcommand{\rada}{0.02249}
\renewcommand{\erada}{0.00031}
\renewcommand{\radb}{0.02224}
\renewcommand{\eradb}{0.00032}
\renewcommand{\lightratio}{0.9343}
\renewcommand{\elightratio}{0.0074}
\renewcommand{\ecc}{0.26725}
\renewcommand{\eecc}{0.00022}
\renewcommand{\omegap}{175.90}
\renewcommand{\eomegap}{0.67}
\renewcommand{\bpri}{1.498}
\renewcommand{\ebpri}{0.019}
\renewcommand{\bsec}{1.557}
\renewcommand{\ebsec}{0.011}
\renewcommand{\rchisq}{1.020}
\renewcommand{\erchisq}{0.069}
\renewcommand{\rmslc}{2.08}
\renewcommand{\rmsprv}{0.36}
\renewcommand{\rmssrv}{0.27}
\renewcommand{\sma}{38.313} 
\renewcommand{\esma}{0.083} 
\renewcommand{\mrat}{0.9595} 
\renewcommand{\emrat}{0.0039} 
\renewcommand{\mpri}{0.3336} 
\renewcommand{\empri}{0.0022} 
\renewcommand{\msec}{0.3200} 
\renewcommand{\emsec}{0.0022} 
\renewcommand{\rpri}{0.862} 
\renewcommand{\erpri}{0.012} 
\renewcommand{\rsec}{0.852} 
\renewcommand{\ersec}{0.013} 
\renewcommand{\logga}{4.090} 
\renewcommand{\elogga}{0.012} 
\renewcommand{\loggb}{4.082} 
\renewcommand{\eloggb}{0.012} 
\renewcommand{\rhoa}{0.521} 
\renewcommand{\erhoa}{0.022} 
\renewcommand{\rhob}{0.517} 
\renewcommand{\erhob}{0.022} 

\clearpage
\LongTables
\begin{deluxetable*}{lrlll}[bt]
\tabletypesize{\scriptsize}
\tablecaption{  System Parameters of UScoCTIO 5 \label{tab:usco5table}}
\tablewidth{0.9\textwidth}
\tablehead{
\colhead{Parameter} & \colhead{Symbol} & \colhead{Value} & \colhead{Units} & \colhead{Source} 
} 

\startdata
Orbital period & $P$ & \period\ $\pm$ \eperiod & days & this work \\
Ephemeris timebase - 2456000 & $T_0$ & \tzero\ $\pm$ \etzero & BJD & this work \\
Surface brightness ratio & $J$ & \sbratio\ $\pm$ \esbratio & & this work \\
Sum of fractional radii & $(R_1+R_2)/a$ & \sumradii\ $\pm$ \esumradii & & this work \\
Ratio of radii & $k$ & \ratradii\ $\pm$ \eratradii &  & this work \\
Orbital inclination & $i$ & \inclination\ $\pm$ \einclination & deg & this work \\
Combined eccentricity, periastron longitude & $e\cos\omega$ & \ecosw\ $\pm$ \eecosw & & this work \\
Combined eccentricity, periastron longitude & $e\sin\omega$ & \esinw\ $\pm$ \eesinw & & this work \\
Primary radial velocity amplitude & $K_1$ & \rvprim\ $\pm$ \ervprim & km s$^{-1}$ & this work \\
Secondary radial velocity amplitude & $K_2$ & \rvsec\ $\pm$ \ervsec & km s$^{-1}$ & this work \\
Systemic radial velocity & $\gamma$ & \rvsys\ $\pm$ \ervsys & km s$^{-1}$ & this work \\
Fractional radius of primary & $R_\text{1}/a$ & \rada $\pm$ \erada & & this work \\
Fractional radius of secondary & $R_\text{2}/a$ & \radb $\pm$ \eradb & & this work \\
Luminosity ratio & $L_2/L_1$ & \lightratio\ $\pm$ \elightratio & & this work \\
Eccentricity & $e$ & \ecc\ $\pm$ \eecc & & this work \\
Periastron longitude & $\omega$ & \omegap\ $\pm$ \eomegap & deg & this work \\
Impact parameter of primary eclipse & $b_1$ & \bpri\ $\pm$ \ebpri & & this work \\
Impact parameter of secondary eclipse & $b_2$ & \bsec\ $\pm$ \ebsec & & this work \\
Orbital semi-major axis & $a$ & \sma\ $\pm$ \esma & \rsun & this work \\
Mass ratio & $q$ & \mrat\ $\pm$ \emrat & & this work \\
Primary mass & $M_1$ & \mpri\ $\pm$ \empri & \msun & this work, fundamental determination \\
Secondary mass & $M_2$ & \msec\ $\pm$ \emsec & \msun & this work, fundamental determination \\
Primary radius & $R_1$ & \rpri\ $\pm$ \erpri & \rsun & this work, fundamental determination \\
Secondary radius & $R_2$ & \rsec\ $\pm$ \ersec & \rsun & this work, fundamental determination \\
Primary surface gravity & $\log g_1$ & \logga\ $\pm$ \elogga & cgs & this work, $M_1$, $R_1$ \\
Secondary surface gravity & $\log g_2$ & \loggb\ $\pm$ \eloggb & cgs & this work, $M_2$, $R_2$ \\
Primary mean density & $\rho_1$ & \rhoa\ $\pm$ \erhoa & $\rho_\odot$ & this work, $M_1$, $R_1$ \\
Secondary mean density & $\rho_2$ & \rhob\ $\pm$ \erhob & $\rho_\odot$ & this work, $M_2$, $R_2$ \\
Reduced chi-squared of light curve fit & $\chi^2_\text{red}$ & \rchisq &  & this work\\
RMS of best fit light curve residuals & & \rmslc & ppt & this work \\
Reduced chi-squared of primary RV fit & $\chi^2_\text{red}$ & 5.58  &  & this work \\
RMS of primary RV residuals & & \rmsprv & km s$^{-1}$ & this work \\
Reduced chi-squared of secondary RV fit & $\chi^2_\text{red}$ & 2.64 &  & this work \\
RMS of secondary RV residuals & & \rmssrv & km s$^{-1}$ & this work
\tablecomments{Best-fit orbital parameters and their uncertainties resulting from  10,000 Monte Carlo simulations with \jktebop. For this analysis we mutually fit our own detrended $K2$ light curve with the radial velocities and spectroscopic flux ratios published in Table 1 of \cite{kraus2015}.}
\enddata

\end{deluxetable*}

\newpage
\bibliography{ms}
\end{document}